\documentclass[useAMS,usenatbib]{mn2e}

\usepackage{graphicx}
\usepackage{amssymb}
\newcommand\msun{{\,M_\odot}}

\newcommand{\unit}[1]{\ensuremath{\, \mathrm{#1}}}
\usepackage{natbib}
\usepackage{cite}
\usepackage{aecompl}
\usepackage[T1]{fontenc}

\newcommand{\mK}{~\mbox{mK}}

\newcommand{\MHz}{~\mbox{MHz}}
\newcommand{\invHz}{~\mbox{Hz}^{-1}}
\newcommand{\erg}{~\mbox{erg}}

\newcommand{\Myr}{~\mbox{Myr}}

\newcommand{\eV}{~\mbox{eV}}
\newcommand{\keV}{~\mbox{keV}}
\newcommand{\Mpc}{~\mbox{Mpc}}

\newcommand{\cMpc}{~\mbox{comoving}~\mbox{Mpc}}
\newcommand{\cpc}{~\mbox{comoving}~\mbox{pc}}

\newcommand{\pc}{~\mbox{pc}}

\newcommand{\ckpc}{~\mbox{comoving}~\mbox{kpc}}

\newcommand{\cmsqi}{~\mbox{cm}^{-2}}
\newcommand{\cmci}{~\mbox{cm}^{-3}}

\newcommand{\kpc}{~\mbox{kpc}}
\newcommand{\K}{~\mbox{K}}

\newcommand{\invyr}{~\mbox{yr}^{-1}}
\newcommand{\invs}{~\mbox{s}^{-1}}

\newcommand{\Msun}{~\mbox{M}_{\odot}}
\newcommand{\Msunyri}{~\mbox{M}_{\odot}~\mbox{yr}^{-1}}

\title[X-ray feedback from HMXBs]{Radiative feedback from high mass X-ray binaries 
on the formation of the first galaxies and early reionization}
\author[Jeon et al.]{Myoungwon Jeon$^{1}$\thanks{E-mail: 
myjeon@astro.as.utexas.edu}, Andreas H. Pawlik$^{2}$, Volker Bromm$^{1}$, Milo\v s Milosavljevi\'{c}$^{1}$\\
$^{1}$Department of Astronomy and Texas Cosmology Center, University of Texas, Austin, TX 78712, USA \\
$^{2}$Max-Planck-Institut f\"ur Astrophysik, Karl-Schwarzschild-Strasse 1, 85748 Garching bei M\"unchen, Germany}

\begin{document}

\date{}

\pagerange{\pageref{firstpage}--\pageref{lastpage}} \pubyear{xxxx}

\maketitle

\label{firstpage}

\begin{abstract}
Recent work suggests that the first generation of stars, the so-called
Population~III (Pop~III), could have formed primarily in binaries or 
as members of small multiple systems. Here we investigate the impact of
X-ray feedback from High-Mass X-ray Binaries (HMXBs) left behind in
stellar binary systems after the primary forms a black hole (BH), 
accreting gas at a high rate from the 
companion, a process that is thought to be favored at the low
metallicities characteristic of high-redshift gas. Thanks to their
large mean free path, X-rays are capable of preionizing and preheating
the gas in the intergalactic medium (IGM) and in haloes long before the
reionization of the Universe is complete, and thus could have strongly
affected the formation of subsequent generations of
stars as well as reionization. We have carried out zoomed
hydrodynamical cosmological simulations of minihaloes, accounting for
the formation of Pop~III stars and their collapse into BHs and HMXBs,
and the associated radiation-hydrodynamic feedback from UV and X-ray photons. 
We find no strong net feedback from
HMXBs on the simulated star formation history. On the other hand, the
preheating of the IGM by HMXBs leads to a strong suppression of
small-scale structures and significantly lowers the recombination rate in the
IGM, thus yielding a net positive feedback on
reionization. We further show that X-ray feedback from HMXBs can augment 
the ionizing feedback from the Pop~III progenitor stars to suppress
gas accretion onto the first BHs, limiting their growth into
supermassive BHs. Finally, we show that X-ray ionization by HMXBs
leaves distinct signatures in the properties of the high-redshift
hydrogen that may be probed in upcoming observations of the redshifted
21~cm spin-flip line.
\end{abstract}


\section{Introduction}
\label{Sec:Intro}
A key question in modern cosmology is to understand how the cosmic
dark ages ended with the emergence of the first sources of light
\citep{Loeb2011, Wiklind2013}. The nature of this transition is
largely driven by the properties of the first stars, also known as 
Population~III (Pop~III). Recent studies of their formation, based on
high-resolution simulations with realistic cosmological initial
conditions, have led to an important revision in our understanding
\citep[e.g.][]{Turk2009, Stacy2010, Stacy2011, Clark2011a, Greif2011, 
Prieto2011, Smith2011, Greif2012, Dopcke2013}. Whereas the previous
standard model had posited that the first stars formed as very
massive, single stars in the centres of dark matter minihaloes
\citep{Bromm2004}, in the revised model they typically emerge as
members of small multiple systems with a range of masses \citep{Bromm2013}.
\par
Although the characteristic mass of the first stars is now predicted
to be smaller than previously thought, of order a few $10$ $M_{\odot}$
instead of $\sim 100$ $M_{\odot}$, the initial stellar mass function
(IMF) is found to be sufficiently broad that stellar masses can still
reach the high values for which black hole (BH) remnants are expected
to form. This is due to the ability to continuously accrete from a
protostellar disk even in the presence of protostellar radiation
 \citep[e.g.][]{Hosokawa2011,Stacy2011,Hirano2013}. These
results open the possibility of Pop~III high-mass X-ray
binaries (HMXBs), where one component is a BH remnant accreting
material from a companion 
\citep[e.g.][]{Mirabel2011, Haiman2011, Xu2013}.
\par
Whether suitable Pop~III binary systems can form is still quite
uncertain \citep[e.g.][]{Saigo2004, Machida2008, Power2009,
  Stacy2012}. The main challenge is to extend the simulations over the
entire accretion phase to ascertain whether the protostars merge or
are ejected by close encounters with other protostars from the natal 
disk. Considering ten different minihaloes for
5000\,yr, recent work by \citet{Stacy2012} found a binary fraction of
$\sim 35\%$ and a distribution of semi-major axes that peaks at $\sim$ 300\,AU.
Depending on the physical properties of binaries, those could be
progenitors of gamma-ray bursts \citep[e.g.][]{Bromm2006,
  Belczynski2007}, or sources of gravitational waves
\citep[e.g.][]{Belczynski2004, Kowalska2012}, possibly detectable by
current and future gravitational wave interferometers.
\par
In the local Universe, the dominant X-ray contribution from
star-forming galaxies comes from HMXBs and ultraluminous X-ray sources
(ULXs) \citep[e.g.,][]{Mineo2012}. HMXBs and ULXs are defined by their
X-ray luminosities as $L_{\rm X}>10^{36} $ erg $\rm s^{-1}$ and $L_{\rm X}>10^{39}
$ erg $\rm s^{-1}$, respectively. Studies of the HMXB population and
its dependence on metallicity, showing that low metallicity favors
HMXB formation \citep[e.g.][]{Majid2004, Dray2006, Soria2007,
  Mapelli2009, Linden2010}, support the importance of HMXBs at high
redshift when the Universe was chemically pristine. This dependence on
metallicity arises in part because in low-metallicity environments
stellar mass-loss via radiatively-driven winds is significantly reduced
\citep{Kudritzki2000}, leading to higher BH remnant masses
\citep{Eldridge2006, Belczynski2010}. However, independent of the
preference for a higher BH mass at low metallicity, the work by
\citet{Linden2010}, employing binary population synthesis
calculations, suggests that metallicity determines the number of
systems undergoing stable mass transfer via Roche lobe overflow
(RLO). This implies that the parameter space of allowed stable
common-envelope pathways for RLO is much wider at low than at 
high metallicity.
\par
The role of radiative feedback by an isolated accreting BH left behind
by a Pop~III star, a so-called miniquasar, has been explored by a
number of authors. Some of them focused on a single source, not
including any later generations of BHs (see, e.g., \citealp{Glover2003}; 
\citealp{Kuhlen2005}; \citealp{Milos2009a};  \citealp{Milos2009b}; 
\citealp{Alvarez2009}; \citealp{Park2011}; \citealp{Venkatesan2011}; \citealp{Wheeler2011}; 
\citealp{Jeon2012}; \citealp{Aykutalp2013}), or semi-analytically on the collective impact of
X-rays from BHs at high redshifts \citep[e.g.][]{Justham2012,
  Tanaka2012, Power2013}.  Given that X-rays can penetrate further
than stellar ionizing photons, their impact can extend to the
distant IGM. In general, X-ray heating may be expected
to provide a strong negative feedback on gas collapse and star formation 
especially near X-ray sources. On the other hand, 
the efficiency of the formation of molecular hydrogen ($\rm H_2$) and hydrogen 
deuteride (HD) and hence molecular cooling may be increased especially in 
distant collapsed structures due to the enhancement in the electron fraction 
by X-ray ionizations, resulting in a positive feedback on star
formation in regions not too close to X-ray sources. Such dependence on distance from a source, however, might
be washed out as the number of X-ray sources increases with time. Thus,
we need to study the collective effect to better understand if the
X-ray feedback on star formation has a positive or negative outcome.
\par
The early establishment of an X-ray background is expected to play an important
role in reionization. X-ray pre-heating might increase the
cosmological Jeans, or filtering, mass.  More time would then be
required to form haloes more massive than the Jeans mass, thus delaying
the process of reionization \citep[see, e.g.][]{Madau2004,
  Ricotti2004, Holley2012, Mesinger2013}. On the other hand, X-ray
heating might reduce small-scale structure, lowering the number
of recombinations, thus in turn accelerating reionization \citep[e.g.][]{
Haiman2001a, Wise2005, Wise2008a, Pawlik2009, Finlator2012, Emberson2013}. 
The net effect is difficult to predict short of treating it with full cosmological simulations. Additionally,
pre-ionization by X-rays may help explain the relatively large value
of the optical depth to reionization measured by the WMAP and Planck
satellites (e.g., \citealp{Ricotti2004}; \citealp{Ahn2012}).
\par
In our preceding paper (\citealp{Jeon2012}), we focused on
investigating the formation of stars and the assembly of the first
galaxies under the radiative feedback from a single isolated 
accreting BH and from a single HMXB, both remnants of a Pop~III star. We found
that with a single accreting BH whose radiative luminosity scales with
the rate of accretion of diffuse halo gas, the gas within the halo hosting the BH can be
photoheated up to $\sim 10^3-10^4$ K, preventing any subsequent star
formation for a few 100\,Myr inside the halo. Distant star formation
outside the host halo, on the other hand, was promoted by the
enhanced formation of $\rm H_2$, which is catalyzed by the
enhanced free electron fraction generated in photoionizations by
the X-rays. The net positive feedback on star formation was however very
mild due to competition between the positive and negative feedback, in agreement with
earlier investigations (\citealp{Machacek2003}; \citealp{Kuhlen2005}).
\par
A stronger effect occurred in the presence of an
HMXB. This was due to the stronger X-ray emission, near the Eddington
limit independent of the conditions in the surrounding interstellar
gas, which photoevaporated the gas and prevented gas infall into the 
halo center, thus establishing a
stronger negative feedback inside the host system. But also the
positive feedback effect was found to be stronger, 
yielding a net increase in the cosmological star formation rate. The
radiation from the HMXB, however, was continuous for
150\,Myr, unrealistically long compared to inferred HMXB lifetimes in the
local Universe, which is on the order of $1-3$\,Myr
\citep{Belczynski2012, Justham2012}. Such an episodic HMXB irradiation
might result in a much weaker effect.  
\par
In this work, we study the radiative feedback from a population of
HMXBs, both locally on the gas in the host halo and globally on the
IGM, using a physically motivated model for HMXBs, and improved
methods to track the accretion of gas and the transport of photons
in the UV and X-ray bands. Computing radiative transfer (RT) of X-ray 
photons is particularly challenging due to the large penetrating power of X-rays that renders each
point in space visible to many sources. The TRAPHIC code
\citep{Pawlik2008, Pawlik2011}, however, allows us to track the
radiation from multiple sources, Pop~III stars and HMXBs, at a 
computational cost that is independent of the number of sources. To
the best of our knowledge, this is the first simulation where the
fully coupled radiation-hydrodynamics with multiple HMXB sources is
taken into account.
\par
This paper is organized as follows. In Section~\ref{Sec:Metho}, we 
describe our simulation methods. In Section~\ref{Sec:Results}, we
present the results of our simulations. We investigate the impact of
HMXBs on the properties of the gas in the IGM and in haloes and
discuss the implications for reionization and black hole growth. In
Section~\ref{Sec:21cm}, we estimate the 21~cm signal of the
high-redshift gas and discuss observational constraints that can be 
derived from the 21 cm signal on the nature of the first ionizing 
sources. Finally, in Section~\ref{Sec:SC}, we
summarize our results. Distances are expressed in physical (i.e., not
comoving) units unless noted otherwise. We will make use of the
species number density fractions with respect to hydrogen $\eta_\alpha
\equiv n_\alpha / n_{\rm H}$, where $\alpha$ labels the chemical
species.

\section{Numerical methodology}
\label{Sec:Metho}
We have carried out cosmological Smoothed Particle Hydrodynamics (SPH)
simulations of the formation of high-redshift minihaloes, following the
chemistry and cooling of the primordial gas, the formation of Pop~III
stars and their collapse into BHs, and the ionization and
photodissociation by the radiation emitted by the stars and accreting
BHs.  
\par
To investigate the feedback by HMXBs we compare two simulations,
which are identical except that in one of them the formation of HMXBs is 
suppressed, and the BHs left behind after the death of the Pop~III 
stars are powered solely by accretion of diffuse halo gas.
\par
In \citet{Jeon2012} we carried out a similar set of simulations of
feedback from Pop~III stars, accreting BHs and HMXBs. The current work
improves on our previous work in several key aspects. Our simulations
are run with an updated version of the SPH code GADGET (last described
in \citealp{Springel2005}), and we employ an improved implementation
of accretion of gas by BHs and a physically motivated model of
HMXBs. The ionizing radiation emitted by Pop~III stars, BHs and HMXBs
is followed using multi-frequency RT, and the spectrum of the radiation emerging from
BHs and HMXBs includes a soft thermal contribution from the 
accretion disk in addition to a power-law high energy
tail. Secondary ionization is accounted for depending on the energies of
the primary photo-electrons. Finally, we track the photodissociation
of $\rm H_2$ and HD by LW
photons both from the Pop~III stars and the accreting BHs, taking into
account local shielding in the gas.
In the following, we provide a detailed description of our
simulations, and refer to our preceding work where appropriate.

\subsection{Gravity, hydrodynamics, and chemistry}
We use the $N$-body/TreePM SPH code GADGET (\citealp{Springel2005}; 
\citealp{Springel2001}; our specific implementation is derived from
that discussed in \citealp{Schaye2010}). Our simulations are
initialized using a snapshot from the earlier zoomed simulation of
\citet{Greif2010}, which was initialized at $z=99$ in a periodic box
of linear size $1 \cMpc$, using $\Lambda$CDM cosmological parameters
and matter density $\Omega_{\rm m}=1-\Omega_{\Lambda}=0.3$, baryon
density $\Omega_{\rm b}=0.04$, present-day Hubble expansion rate $H_0
= 70\unit{km\, s^{-1} Mpc^{-1}}$, spectral index $n_{\rm s}=1.0$, and 
normalization $\sigma_8=0.9$, which is higher than the value measured 
by the WMAP and Planck satellites (e.g., \citealp{Komatsu2011};
\citealp{planck2013}). The choice of a high value of $\sigma_8$ might 
accelerate structure formation in our simulations. However, in this 
paper, we focus on the effects of radiative feedback by comparing individual 
haloes, and this comparison is insensitive to the variation in the $\sigma_8$ parameter. Employing consecutive levels of refinement, the
masses of dark matter (DM) and SPH particles in the highest resolution
region with an approximate extent of $300 \ckpc$ are $m_{\rm DM}
\approx 33\msun$ and $m_{\rm SPH} \approx 5 \msun$, respectively.
\par
We start the simulations at $z \approx 30$, corresponding to the time 
just before the first Pop~III star is formed, and terminate them at
$z\approx 18$. We adopt a Plummer-equivalent gravitational softening
length $\epsilon_{\rm soft} = 70 \cpc$ for both dark matter and
baryonic particles. SPH quantities are estimated by averaging inside a
sphere containing $N_{\rm ngb} = 48$ neighbors and adopting the
entropy conserving formulation of SPH (\citealp{Springel2002}). The
SPH kernel, i.e., the radius of this sphere, is prevented from falling
below $10^{-3} \epsilon_{\rm soft}$. The gas particle mass
defines a baryonic mass resolution $M_{\rm res} \equiv N_{\rm
  ngb} m_{\rm SPH}\approx 240\msun$. The Jeans mass in primordial gas, 
in which molecular hydrogen cooling imprints a condition for primary 
star formation with a characteristic density of $n_{\rm H}\simeq 10^4 
\unit{cm^{-3}}$ and a temperature of $\sim 200$ K (\citealp{Abel2000}; 
\citealp{Bromm2002}), is thus marginally resolved, and the \citet{Bate1997} 
criterion to avoid artificial fragmentation is marginally satisfied.
\par
We use the same primordial chemistry and cooling network as in \citet{Greif2010},
where all relevant cooling mechanisms, i.e., by H and He collisional
ionization, excitation and recombination cooling, bremsstrahlung,
inverse Compton cooling, and collisional excitation cooling of $\rm
H_{\rm 2}$ and HD, are taken into account. The radiative cooling by 
$\rm H_{\rm 2}$ is computed accounting for collisional 
excitations by protons and electrons, which is important in 
gas with a significant fractional ionization (\citealp{GloverAbel2008}). 
The code self-consistently solves the rate
equations for the abundances of $\rm H, H^{+}, H^{-}, H_{2}, H^{+}_2 ,
He, He^{+}, He^{++},$ and $\rm e^{-}$, as well as for the three deuterium
species $\rm D, D^{+}$, and HD, taking into account photoionization, 
secondary ionization, and photodissociation as described below. 
We set the hydrogen mass fraction to $X=0.76$.
\par
\subsection{Sink particles}
\label{Sec:sink}
Upon reaching hydrogen number densities $n_{\rm H, max}=10^4 \cmci$,
gas particles are converted into collisionless sink particles. Each
newly formed sink particle accretes a random subset of the $N_{\rm
  ngb} = 48$ neighboring gas particles residing in the SPH smoothing
kernel, which defines the effective accretion radius, $r_{\rm acc}$,
of the progenitor gas particle, until its mass has exceeded
$100\Msun$.  The initial mass of sinks is thus $M_{\rm sink} \approx
100 \msun$, which is appropriate given the adopted star formation
model that we describe below.
\par
\par
Sink particles can grow in mass by accreting gas inside the
sphere with radius $r_{\rm acc}$. Sink particles that approach each other on
scales below $1 \pc$, which is similar to the baryonic resolution scale 
$l_{\rm res} \equiv [( 3X M_{\rm res})/(4 \pi n_{\rm H, max}
  m_{\rm H})]^{1/3} \approx 0.5 \pc$, are merged. The 
position and velocity of the sink particle resulting from the merger
are set to the mass-weighted average position and velocity of the
sink particles before merging.
\par
In the following, we identify the sink particles with the sites of Pop~III 
star formation.

\subsection{Pop~III binaries}
Our goal is to investigate the feedback from HMXBs which may follow the 
formation of stellar binaries. We therefore assume for simplicity, 
but consistent with recent work on fragmentation of 
primordial gas (e.g., \citealp{Turk2009, Stacy2010, Stacy2011, Clark2011a, Greif2011, 
Prieto2011, Smith2011, Greif2012, Dopcke2013}), that all Pop~III stars form in binaries. Fragmentation and the formation
of binaries are treated using a sub-resolution model, assuming each
sink is host to a Pop~III binary system. To limit the instantaneous
binary formation rate to one system per minihalo, we prevent
the creation of additional sink particles for $t_{\rm sh} =1$\,Myr in
spheres with radius $r_{\rm sh}=30 \pc$ centred on each newly
created sink particle. This corresponds to the time needed for
photoheating by a Pop~III star in a minihalo to suppress the gas density inside the spheres,
assuming the resulting hydrodynamic shock propagates at $v_{\rm sh}
\approx 30$ $\rm km\, s^{-1}$ (e.g., \citealp{Shu2002}; \citealp{Whalen2004}; \citealp{Alvarez2006}). 
\par
In practice, the adopted recipe guarantees there can be at most a single active Pop~III
binary per minihalo at any given time. The properties of the binary
system are assigned assuming that the primary dominates
the mass of the system.  Thus, the stellar properties of the binary
are set to that of an isolated Pop~III star of mass 100$\msun$. Such
a star shines for $2.7 \Myr$ emitting a black body spectrum with
temperature $T_{\rm BB}=10^5$ K, and has a hydrogen-ionizing photon luminosity of $9.14 \times 10^{49} \invs$ 
and a photodissociating luminosity of 
$1.1 \times 10^{49} \invs$ in the LW band (e.g., \citealp{Bromm2001a}; 
\citealp{Schaerer2003}). 

\subsection{Black hole miniquasars}
\label{Sec:xray}
The fate of Pop~III stars is determined by their initial masses 
\citep{Heger2003}. Here, we assume that the more massive
Pop~III companion in each binary system collapses into a BH with mass
$100 \Msun$. Recent studies, in which the effects of rotation of Pop~III
stars is taken into account, have suggested that stars with masses
$\gtrsim 60-240 \Msun$ may explode as pair instability supernovae
without leaving a BH behind (e.g., \citealp{Chatzopoulos2012},
\citealp{Yoon2012}). However, our conclusions are not sensitive to the
adopted initial mass of the BHs.
\par
The BH then grows by accretion of the surrounding baryons at a rate \citep{Bondi1944} 
\begin{equation}
\dot{M}_{\rm acc} = \frac{4 \pi G^2 M_{\rm BH}^2 \rho}{(c_{\rm s}^2 + v_{\rm
    rel}^2)^{3/2}},
\end{equation}
where $c_{\rm s}$ is the sound speed, $\rho$ the gas density, and 
$v_{\rm rel}$ the velocity of the BH relative to the gas. We assume that a 
fraction $\epsilon = 0.1$ of the accreted mass is converted into
ionizing radiation, as is appropriate for 
radiatively efficient accretion onto a Schwarzschild
BH \citep{Shakura1973}. The BH thus grows in mass according to $\dot{M}_{\rm BH} =
(1-\epsilon) \dot{M}_{\rm acc}$. If the black hole is accreting at a high rate, 
it is referred to as a miniquasar (e.g., \citealp{Kuhlen2005}).
\par
The density, sound speed and relative velocity needed to evaluate the
accretion rate are estimated by averaging the properties of the gas
particles inside the sphere with radius $r_{\rm acc}$ centred on the
sink particle hosting the BH, consistent with the SPH formulation.
For reference, the Bondi-Hoyle radius is  (e.g., \citealp{Edgar2004}) $r_{\rm Bondi} \equiv \mu
m_{\rm H} G M_{\rm BH} / (k_{\rm B} T) \approx 0.6 \pc (\mu / 1.2)
(M_{\rm BH} / 100 \Msun)/(T / 100 \K)$, where $\mu$ is the mean molecular 
weight.
\par 
The mass of the sink particle $M_{\rm sink}$ hosting the BH closely
tracks the mass $M_{\rm BH}$ of the BH. We achieve this by letting the
sink particles swallow neighboring gas particles, randomly chosen
inside the sphere with radius $r_{\rm acc}$, as long as the condition
$M_{\rm BH}-M_{\rm sink} > m_{\rm SPH}$ is satisfied (e.g.,
\citealp{Springel2005b}; \citealp{Dimatteo2005}; \citealp{Sijacki2007}; \citealp{booth2009}).  This
improves our implementation of BH growth in \citet{Jeon2012}, in
which the BHs and sink particles grew independently. 
\par
We compute the ionizing luminosities of BHs accreting diffuse gas by normalizing the total luminosity according to
\begin{eqnarray}
L_{\rm BH} &\equiv& \int_{0}^{10 {\rm keV}/h_{\rm P}} {L_{\nu} d\nu} = \frac{\epsilon}{1-\epsilon} \dot{M}_{\rm BH} c^2 \\
&=& 6.4 \times 10^{37} {\rm erg \hspace{0.1cm} s^{-1}} \frac{\dot{M}_{\rm BH}}{10^{-8} \Msunyri},
\end{eqnarray}
where $h_{\rm P}$ is Planck's constant. We further assume that the spectral energy distribution of the emerging
radiation is that of a thermal multi-color disk (MCD) at frequencies
lower than $\nu _{\rm NT} \equiv 0.2 \keV/h_{\rm P}$, and that of a non-thermal
power-law component at higher frequencies, consistent with
observations at low redshifts (see Figure~1 in
\citealp{Jeon2012}; e.g., \citealp{Mitsuda1984}; \citealp{Miller2003};
\citealp{Madau2004}; \citealp{Kuhlen2005}).  
\par
The spectrum of the radiation emitted by the
multi-color disk with gas temperature profile $T=T_{\rm in} (r/r_{\rm
  in})^{-p}$, where $r$ is the distance between the BH and the
location inside the disk at which the radiation emerges and $r_{\rm
  in}$ is the inner disk radius, is (e.g., \citealp{Pringle1981};
\citealp{Mitsuda1984})
\begin{equation}
J_{\rm MCD}(\nu) \propto \left( \frac{k_{\rm B} T_{\rm in}}{h_{\rm P}\nu} \right)^{2/p} \nu^3 \int_{x_{\rm in}}^{x_{\rm out}} {dx \frac{x^{2/p-1}}{e^{x}-1}}, 
\end{equation}
where $T_{\rm in}$ is the gas temperature at $r=r_{\rm in}$, $ x_{\rm
  in} = h_{\rm P} \nu / (k_{\rm B}T_{\rm in})$, and $x_{\rm out} =
h_{\rm P} \nu / (k_{\rm B}T_{\rm in}) (r_{\rm out}/r_{\rm in})^p$. The
inner radius $r_{\rm in}$ is set to the radius of the last stable
Keplerian orbit around the BH, which for non-rotating BHs is $r_{\rm
  in}=3$$r_{\rm s}$, where $r_{\rm s}=2GM_{\rm BH}/c^2$ is the
Schwarzschild radius, and the outer radius is set to $r_{\rm out}=10^4
r_{\rm in}$.  We adopt $p=3/4$, appropriate for viscous standard disks
(\citealp{Shakura1973}; \citealp{Pringle1981}; \citealp{Mitsuda1984};
\citealp{Kato1998}) and set $T_{\rm in} = [3 GM_{\rm BH} \dot{M}_{\rm 
acc} /(8 \pi \sigma_{\rm SB} r_{\rm in}^3) ]^{1/4}$ (e.g.,
\citealp{Pringle1981}; \citealp{Kato1998}), where $\sigma_{\rm SB}$ is
the Stefan-Boltzmann constant. 
\par
The radiation spectrum of the power-law component is
\begin{equation}
J_{\rm NT}(\nu) \propto \left( \frac{\nu}{\nu_{\rm NT}}
\right)^{-\beta},
\end{equation} 
where we set $\beta=1$ (e.g., \citealp{Kuhlen2005}). Each of the components 
of the two-component spectrum is normalized by assuming that a half of the 
total luminosity emerges from each of the MCD and the power law component 
(e.g., \citealp{Kuhlen2005}).
\par
The two-component spectrum implies an ionizing luminosity in the range $13.6
\eV-10 \keV$ of $3.1 \times 10^{35} - 3.1\times10^{39} \erg \invs$ for a BH 
with mass $100 \Msun$ accreting gas at rates $10^{-10}-10^{-6}
\Msunyri$. We also compute the luminosity of the accreting BH in the
LW bands by integrating the spectrum over the
range of energies $11.2-13.6 \eV$. The LW luminosities of a BH with
mass $100 \Msun$ accreting gas at rates encountered in our simulations 
$10^{-10}-10^{-6} \Msunyri$ are $2.3 \times 10^{33}-9.6 
\times10^{36} \rm erg \invs$.

\subsection{High-mass X-ray binaries}
\label{Sec:HMXB}
In one of the two simulations presented here, for a subset of the relic BHs, 
the primary accretes gas for a brief time 
directly from the low-mass stellar companion, giving rise to a HMXB.
\par
The fraction of high-redshift stellar binaries evolving into HMXBs is
uncertain. \citet{Power2009} investigated the contribution of the HMXB
population in globular clusters to the high-redshift X-ray
background. They used Monte Carlo models of globular clusters
containing $10^6$ stars and exploring a range of stellar IMFs
and initial binary orbital parameters. They found that, assuming 
a \citet{Kroupa2001} IMF, up to 30 per cent of the initial binary systems 
may avoid dissociation after the primary star undergoes a supernova explosion,
thus possibly leading to an HMXB. 
\par
Motivated by the results of \citet{Power2009}, we select every
third binary to evolve into an HMXB immediately after the primary turned 
into a BH with mass $100 \Msun$. We adopt a duration 
of the HMXB phase of $2 \Myr$, approximately corresponding to the remaining 
main sequence lifetime of the donor star from which the BH 
accretes, consistent with local observations \citep[e.g.,][]{Belczynski2012}.  We assume that the HMXB has a 
luminosity equal to the Eddington luminosity, i.e.,
\begin{eqnarray}
L_{\rm HMXB} &\equiv &  \int_{0}^{10 {\rm keV}/h_{\rm P}} {L_{\nu} d\nu} = L_{\rm Edd} \\
&=& 1.4 \times 10^{40} {\rm erg \hspace{0.1cm} s^{-1}}
\left( \frac{M_{\rm BH}}{100 \msun} \right),
\end{eqnarray}
corresponding to accretion of gas from the companion at a rate
$2.2 \times 10^{-6} \Msunyri (M_{\rm BH} / 100 M_{\odot})$ and a total
accreted mass at the end of the HMXB phase of about $4.4 M_{\odot}$.  
The adopted accretion luminosity is consistent with the
inferred luminosities of ULXs
in local surveys \citep[e.g.,][]{Grimm2003}. We employ the
two-component spectrum of the previous section used to describe the
emission of radiation by BHs accreting diffuse halo gas.  Hence,
assuming accretion at the Eddington rate and a BH mass of $100 M_{\odot}$,
the ionizing photon luminosity of a HMXB is $3.3 \times 10^{49}\invs$,
and its LW luminosity is $4.7\times 10^{47} \invs$. The role of the
donor star is limited to fueling the accreting BH, i.e., we ignore
effects of its evolution, such as the emission of ionizing
radiation or the explosion in a SN.

\begin{figure*}
  \includegraphics[width=175mm]{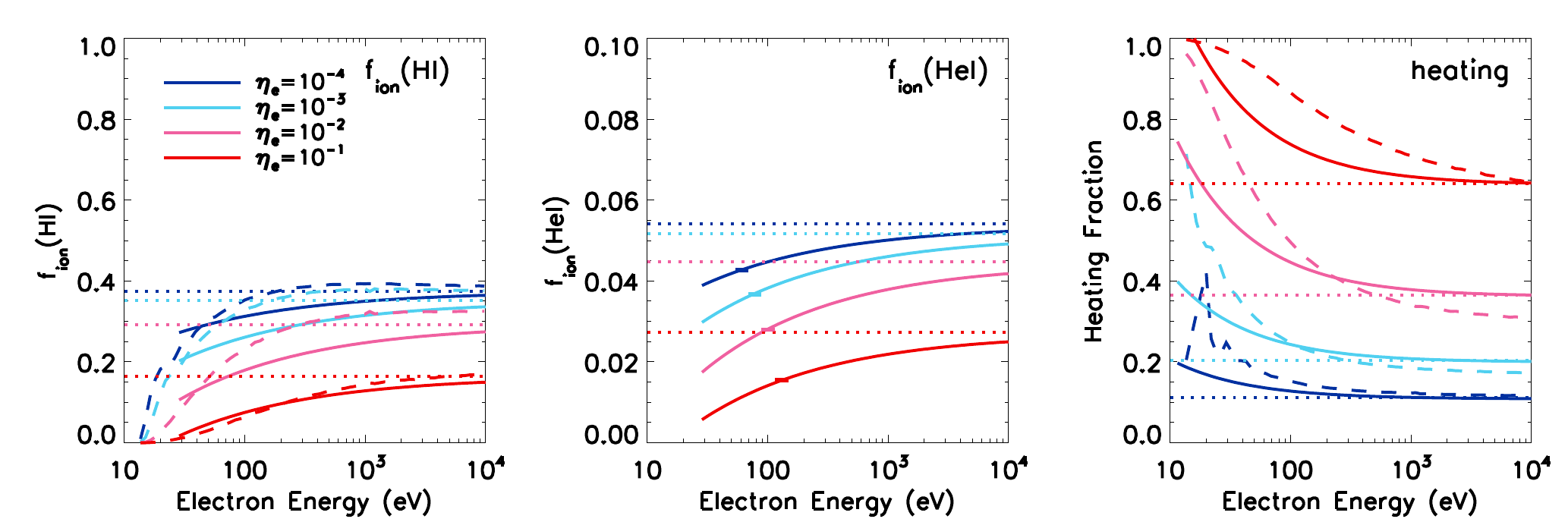}
    \caption{Fraction of the energy of the
      primary electron that is used in HI secondary ionization (left
      panel), HeI secondary ionization (middle panel) and heating
      (right panel). Different colors show these fractions assuming
      different electron fractions. The solid curves show the fits
      from \citet{Ricotti2002} used in the current work. The dashed
      curves show the more recent fits by \citet{Furlanetto2010} for
      comparison. The horizontal dotted lines show the high-energy
      limits from \citet{Shull1985}, which we adopted independent of
      energy in \citet{Jeon2012}.  The short bars in the middle panel
      indicate the energies of photons with mean free path equal to
      the physical size of the high-resolution region of comoving extent 
      $300 \kpc$ in which we follow the transfer of ionizing radiation from 
      $z = 30-18$. \label{Fig:secion}}
\end{figure*}

\subsection{Hydrogen and helium ionizing radiative transfer and X-ray secondary ionization}
\label{Sec:hydro}
We use the RT code TRAPHIC to transport ionizing
photons \citep{Pawlik2008,Pawlik2011}. TRAPHIC solves the
time-dependent RT equation in SPH simulations by tracing photon
packets emitted by source particles through the simulation box in a photon-conserving manner. 
The photon packets are transported directly on the spatially adaptive set of 
SPH particles and hence the RT exploits the full dynamic range of the hydrodynamical
simulations. A directed transport of the photon packets radially away from
the sources is accomplished despite the irregular distribution of SPH
particles by guiding the photon packets inside cones. A photon packet
merging technique renders the computational cost of the RT independent
of the number of radiation sources.  In the following, we provide a
brief overview of TRAPHIC in order to motivate the meaning of the
numerical parameters of the RT specified below. The reader is referred
to the descriptions in \citet{Pawlik2008}, \citet{Pawlik2011} and
\citet{Pawlik2013} for details.  The specific version employed here is
identical to that used in \citet{Pawlik2013}, except for the treatment
of secondary ionization, which is included only here and described
below.

\subsubsection{Basic principles}

The transport of radiation starts with the emission of photon packets
by source particles (here, sinks) in $N_{\rm EC}$ tessellating {\it
  emission cones}. The photons in each
photon packet are distributed among the subset of the $\tilde{N}_{\rm
  ngb} \lesssim N_{\rm ngb}$ neighboring SPH particles residing in the cones.  In
cones containing zero neighbors, an additional, so-called virtual particle is
inserted to which the photon packet is then assigned. Sources emit
photons using emission time steps $\Delta t_{\rm em}$, in between which
the orientation of the cones is randomly rotated to increase the
sampling of the volume with photons. The spectrum of the emitted
radiation is discretized using $N_{\nu}$ frequency bins. Each photon
packet carries photons from one of these bins; therefore the
number of photon packets emitted per emission time step is
$N_{\nu}\times N_{\rm EC}$.
\par
The newly emitted photons are assigned a propagation direction
parallel to the central axis of the associated emission cone, and,
together with any other photons already present in the simulation box,
are then propagated further to the {\it downstream} neighbors of the SPH
particle at which they reside. A particle is a downstream neighbor if
it is among the $\tilde{N}_{\rm ngb}$ neighboring gas particles and
resides in the regular {\it transmission cone} centred on the
propagation direction and subtending a solid angle of $4 \pi/N_{\rm
  TC}$. The parameter $N_{\rm TC}$ hence defines the angular
resolution of the RT. If there is no downstream neighbor inside a
transmission cone, like for emission, a virtual particle is created to which the photons
are then propagated. The transmission cones confine the propagation of
photons to the solid angle in which they were originally emitted. The
transport of photons, which occurs at the user-specified speed
$\tilde{c}$, is discretized using RT time steps $\Delta t_{\rm r}$.
\par
A given SPH particle can receive multiple photon packets within the same
RT time step $\Delta t_{\rm r}$. These photon packets are grouped according to their propagation
directions using a set of $N_{\rm RC}$ tessellating {\it reception
  cones}. Photon packets whose propagation directions fall in the same
reception cone are merged and replaced by a single new photon
packet. Each reception cone subtends a solid angle $4\pi / N_{\rm
  RC}$, and hence the parameter $N_{\rm RC}$ determines the angular
resolution of the merging. The merging limits the maximum number of
photon packets in the simulation box to $N_{\rm RC} \times N_{\nu}
\times N_{\rm SPH}$, where $N_{\rm SPH}$ is the number of SPH
particles, and renders the computation time independent of the number
of sources.
\par
Photons are absorbed as they propagate through the gas from SPH
particles to their neighbors depending on the optical depth between the
two neighboring particles, respecting photon conservation (\citealp{Abel1999}; \citealp{Mellema2006}). 
The absorption of photons within each frequency bin is
treated in the grey approximation using photoionization cross sections
\begin{equation}
\langle \sigma_{\alpha} \rangle_{\nu} \equiv \int_{\nu_{\rm l}}^{\nu_{\rm h}} {d\nu \frac{4 \pi J_{\nu}(\nu)}{h_{\rm P}\nu} 
						  \sigma_{\alpha}(\nu)} 
						  \times
						   \left[ \int_{\nu_{\rm l}}^{\nu_{\rm h}} 
						  {d\nu \frac{4 \pi J_{\nu}(\nu)}{h_{\rm P} \nu} }  \right]^{-1},
\end{equation}
where $\alpha \in \{\rm HI, HeI, HeII\}$, $J_{\nu}(\nu)$ is the
spectrum, and $\nu_{\rm l}$ and $\nu_{\rm h}$ are the low and high energy limits of frequency 
bin\footnote{For brevity, we use the symbol $\nu$ both to number the frequency bin and to denote the frequency.} $\nu$. 
The number of absorbed photons determines the 
photoionization rate $\Gamma_{\gamma \alpha, \nu}$ of species $\alpha$ in the given 
frequency bin $\nu$ defined by \citep[e.g.,][]{Osterbrock2006}
\begin{equation}
\Gamma_{\gamma \alpha, \nu} = \langle \sigma_{\alpha} \rangle_{\nu} \int_{\nu_{\rm l}}^{\nu_{\rm h}} {d\nu \frac{4 \pi J_{\nu} (\nu)}{h_{\rm P} \nu} }.
\end{equation}
The photoionization rate implies a photoheating rate given by 
$\mathcal{E}_{\gamma \alpha, \nu} = \langle \varepsilon_{\alpha} \rangle_{\nu} \Gamma_{\gamma \alpha, \nu}$, where
\begin{eqnarray}
\langle \varepsilon _{\alpha} \rangle_{\nu} &=& \left[  \int_{\nu_{\rm l}}^{\nu_{\rm h}} {d\nu \frac{4 \pi J_{\nu} (\nu)}{h_{\rm P} \nu} 
						   \sigma_{\alpha} (\nu) (h_{\rm P}\nu - h_{\rm P}\nu_{\alpha})} \right] \nonumber \\
						  &\times& 
						   \left[ \int_{\nu_{\rm l}}^{\nu_{\rm h}} {d\nu \frac{4 \pi J_{\nu}(\nu)}{h_{\rm P} \nu} }  
						   \sigma_{\alpha} (\nu) \right]^{-1}						  
\end{eqnarray}
is the grey excess energy of frequency bin $\nu$, and $h\nu_\alpha$ is the photoionization threshold energy of species $\alpha$. For reference, 
$h\nu_{\rm HI} = 13.6\eV$, $h \nu_{\rm HeI} = 24.6\eV$,  and $h \nu_{\rm HeII} = 54.4\eV$.

\subsubsection{Secondary ionizations}

The photoionization and photoheating rates are passed to the chemistry solver described above, which
updates the abundances of hydrogen and helium that evolve according to the following set of equations (e.g., \citealp{Kuhlen2005}), 

\begin{eqnarray}
\frac{d\eta_{\rm HI}}{dt} &=& \alpha_{\rm HII} n_{\rm e} \eta_{\rm HII} - \eta_{\rm HI} \Gamma_{\rm \gamma HI} - \eta_{\rm HI} \Gamma_{\rm eHI} n_{\rm e} \nonumber \\
		                        &-&f_{\rm ion, HI}  \frac{ ( \eta_{\rm HI} \mathcal{E}_{\rm HI} +\eta_{\rm HeI} \mathcal{E}_{\rm HeI})} { 13.6 \rm eV},
\end{eqnarray}

\begin{eqnarray}
\frac{d\eta_{\rm HeI}}{dt} &=& \alpha_{\rm HeII} n_{\rm e} \eta_{\rm HeII} - \eta_{\rm HeI} \Gamma_{\rm \gamma HeI} - \eta_{\rm HeI} \Gamma_{\rm eHeI} n_{\rm e} \nonumber \\
		                        &-&f_{\rm ion, HeI}  \frac{ (\eta_{\rm HI} \mathcal{E}_{\rm HI} + \eta_{\rm HeI} \mathcal{E}_{\rm HeI})} { 24.6 \rm eV},
\end{eqnarray}

\begin{equation}
\frac{d\eta_{\rm HeIII}}{dt} = \eta_{\rm HeII} \Gamma_{\rm \gamma HeII} + \eta_{\rm HeII} \Gamma_{\rm eHeII} n_{\rm e}  -\alpha_{\rm HeIII} n_{\rm e} \eta_{\rm HeIII},		                     
\end{equation}
respecting the constraints 
\begin{eqnarray}
\eta_{\rm HI} + \eta_{\rm HII} &=& 1, \\
\eta_{\rm HeI} + \eta_{\rm HeII} + \eta_{\rm HeIII} &=& \eta_{\rm He}, \\
\eta_{\rm HII} + \eta_{\rm HeII} + 2 \eta_{\rm HeIII} &=& \eta_{\rm e},
\end{eqnarray}
where $\Gamma_{{\rm e}\alpha}$ and $\alpha_\alpha$ are the collisional ionization and
recombination rate coefficients for species $\alpha$. The helium abundance
is defined as $\eta_{\rm He} = n_{\rm He}/n_{\rm H} = X_{\rm
  He}(m_{\rm H}/m_{\rm He})/ (1-X_{\rm He})$ and $m_{\rm H}$ and $m_{\rm He} = 4 m_{\rm H}$ are 
the masses of the hydrogen and helium atoms. 
\par
The factors $f_{\rm ion, HI}$ and $f_{\rm ion, HeI}$ describe
the secondary ionizations of HI and HeI by energetic electrons. We 
use the ionization and energy dependent fits by \citet{Ricotti2002}
to the results of \citet{Shull1985}, i.e., 

\begin{eqnarray}
f_{\rm ion, HI} \approx &-&0.69 \left( \frac{28 \rm eV}{E}\right)^{0.4} \eta_{\rm e}^{0.2} (1-\eta_{\rm e}^{0.38})^2 \nonumber \\
		          &+&0.39 (1 - \eta_{\rm e}^{0.41})^{1.76}, 
\end{eqnarray}

\begin{eqnarray}
f_{\rm ion, HeI} \approx &-&0.098 \left( \frac{28 \rm eV}{E}\right)^{0.4} \eta_{\rm e}^{0.2} (1-\eta_{\rm e}^{0.38})^2 \nonumber \\
		          &+&0.55 (1 - \eta_{\rm e}^{0.46})^{1.67},
\end{eqnarray}
for $ E > 28 \eV$ and $f_{\rm ion, HI}= f_{\rm ion, HeI} = 0 $ for $E
< 28 \eV$, where $E = h_{\rm P}(\nu - \nu_{\alpha})$ is the energy of
the primary electron resulting from photoionizations of species
$\alpha$ by photons with energy $h_{\rm P}\nu$. We follow \citet{Ricotti2002}
and neglect secondary ionization and excitation of
HeII. \citet{Ricotti2002} also provided fits to the factors $f_{\rm
  heat}$ by which the rate at which the gas is photoheated must be
reduced to account for the energy lost in secondary ionizations and
excitations,
\begin{eqnarray}
f_{\rm heat} &\approx& 3.9811 \left( \frac{11 \rm eV}{E}\right)^{0.7} \eta_{\rm e}^{0.4} (1-\eta_{\rm e}^{0.34})^2 \nonumber \\
		            &+& \left[ 1- (1 - \eta_{\rm e}^{0.27})^{1.32} \right],
\end{eqnarray}
for $E > 11 \eV$ and $f_{\rm heat}= 1$ for $E < 11
\eV$. Figure~\ref{Fig:secion} compares the energy dependent fits by
\citet{Ricotti2002} at a range of electron fractions with the
\citet{Shull1985} high-energy limit that we employed in
\citet{Jeon2012} independent of electron energy. The figure also shows
the results of the more recent work by \citet{Furlanetto2010}. 

\begin{figure*}
   \includegraphics[width=110mm]{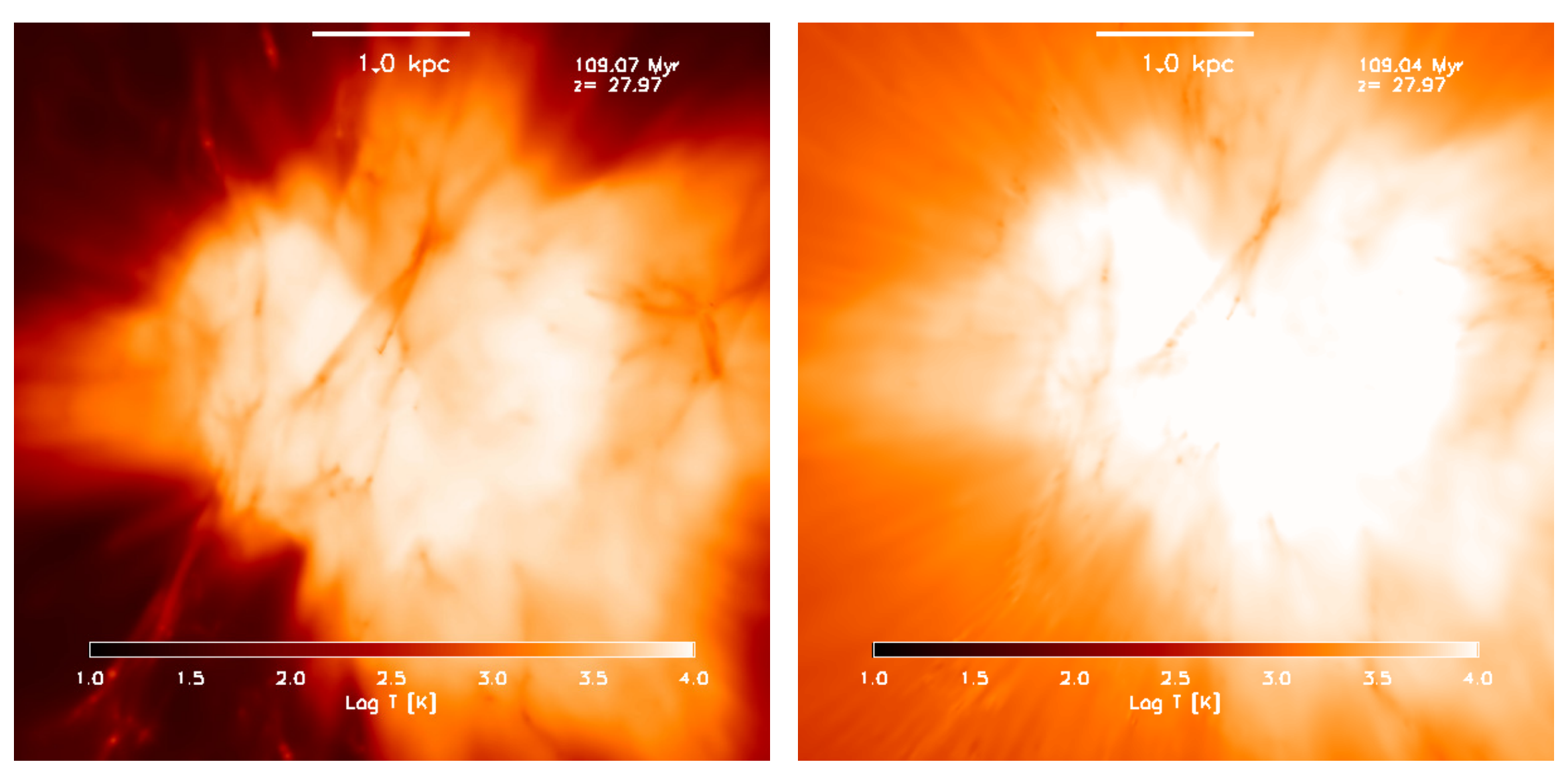}
   \caption{Temperatures, averaged along the line of sight, in a
     cubical cutout of linear extent $150 \ckpc$ centred on the first
     sink particle $2 \Myr$ after the formation of the first BH, in
     the simulation in which the BH grew by accreting diffuse
     halo gas ({\it left}), and in the simulation in which the BH grew
     by accretion from the stellar companion in a HMXB ({\it
       right}). In the former case, because of the low Bondi-Hoyle
     accretion rates, the ionizing luminosities are negligible, and
     the butterfly-shaped photoheated relic HII region created by the
     Pop~III progenitor star dominates the image. In the latter case,
     on the other hand, the large X-ray luminosity of the HMXB implies
     a strong preheating of the IGM outside of the stellar relic HII
     region.  \label{Fig:FirstStarSlices}}
\end{figure*}

\subsubsection{Numerical parameter choices}

In the simulations presented in this work, we set $\tilde{N}_{\rm ngb}
= 32$, and choose an angular resolution of the transport of $N_{\rm
  TC} = 128$ and of the merging of $N_{\rm RC} = 8$. These choices
have been made after carrying out a set of comparison runs and
testing for convergence.  Sources emit photons into $N_{\rm EC} = 128$
directions using emission time steps $\Delta t_{\rm em} = \min
(10^{-3}\Myr, \Delta t_{\rm r})$. Photons are transported at a speed $\tilde{c} = 0.1c$, 
where $c$ is the speed of light, using time steps of size $\Delta t_{r} = \min 
(10^{-2}\Myr, \Delta t_{\rm hydro})$, where $\Delta t_{\rm hydro}$ is the 
smallest GADGET particle time step. The reduction of
the speed of light by a factor 10 does not noticeably affect the
outcome of our simulations, but reduces the simulation cost. We use
four unevenly spaced frequency bins with bounding energies located at [13.6, 24.6, 54.4, 400,
  $10^4$] eV. Test runs using a larger number of frequency bins
yielded nearly indistinguishable results, consistent with the
discussion in \citet{Mirocha2012}.
\par
We use the fits to the frequency-dependent photoionization cross
sections by \citet{Verner1996}, and simplify the computation of the
grey photoionization cross sections $\langle \sigma_{\alpha}
\rangle_{\nu}$ and of the grey excess energies $\langle \varepsilon_{\alpha}
\rangle_{\nu}$ by adopting a fixed spectrum independent of the source
from which the photons in frequency bin $\nu$ originate. To improve on
this approximation, we compute the grey averages using the spectrum of 
the source formed last, assuming accretion at the Eddington rate and a BH mass of 
$100 \Msun$ in the case of X-ray
sources. Tests in which we varied the adopted spectrum over
the range of typical spectra of the sources suggested 
no significant impact of this approximation on our results. An
accurate treatment in which individual grey cross-sections and excess
energies are assigned to each photon packet will be pursued in future
work. For simplicity, we do not compute grey averages of the energy dependent 
functions  $f_{\rm ion, HI}, f_{\rm ion, HeI}$, and $f_{\rm heat}$ describing 
secondary ionization, but evaluate them at the characteristic frequency determined 
by averaging the frequencies inside the bins weighted by the spectrum of the source. 
\par
To reduce the computational expense of the simulations we do not trace
photons emitted by X-ray sources with accretion rates below
$\dot{M}_{\rm BH}<10^{-10} \Msunyri$. Tests showed that the radiative feedback
from BHs with such low accretion rates is negligible, consistent with
our discussion in \citet{Jeon2012}. We also do not follow the propagation
of photons outside the high-resolution region.

\subsection{Photodissociation of $\rm H_2$ and HD}
\label{Sec:LW}
The molecules $\rm H_2$ and HD are photodissociated by LW photons with
energies in the range $11.2-13.6 \eV$. We follow previous works (e.g.,
\citealp{Wise2008a}; \citealp{Wolcott2011b};
\citealp{SSC2012}; \citealp{Johnson2013}; \citealp{Agarwal2012}) and compute the
photodissociation rates in the optically thin limit and apply a
shielding correction to approximate RT effects. The photodissociation
rate of $\rm H_2$ is thus (\citealp{Abel1997})
\begin{equation}
k_{\rm H_2} = 1.1 \times 10^8 \invs  \frac{f_{\rm shield, H_2} F_{\rm LW}}{\erg \invHz \invs \cmsqi}
\end{equation}
and the photodissociation rate of HD is (\citealp{Wolcott2011})
\begin{equation} 
k_{\rm HD} = 1.1  \times 10^8 \invs   \frac{ f_{\rm shield, HD} f_{\rm shield, H_2, HD} F_{\rm LW}}{\erg \invHz \invs \cmsqi},
\end{equation}
where $F_{\rm LW}=4 \pi J_{\rm LW}$ is the flux in the LW bands
computed in the optically thin limit, i.e., ignoring any absorption of
the LW photons by the gas, and $J_{\rm LW}$ the corresponding intensity. 
\par
The dimensionless factors $f_{\rm shield, H_2}$, $f_{\rm shield, HD}$, and $f_{\rm
  shield, H_2, HD}$ describe the local attenuation of the flux by
self-shielding of $\rm H_2$, self-shielding of HD, and shielding of HD by $\rm H_2$. 
We compute the self-shielding factors $f_{\rm shield, H_2}$ and $f_{\rm shield, HD}$ 
using (\citealp{Wolcott2011b})
\begin{eqnarray}
f_{\rm shield} (N, T) = \frac{0.965}{(1+x/b_{5})^\alpha}+ \frac{0.035}{(1+x)^{0.5}} \nonumber \\
             \times \exp{[-8.5 \times 10^{-4}(1+x)^{0.5}]},
\end{eqnarray}
where $x \equiv N_i /5\times10^{14}\rm cm^{-2} $, $N_i$ is the column
density of the molecule species, $b_5\equiv b/10^5 \rm cm$ $\rm
s^{-1}$, $b \equiv \sqrt{2k_{\rm B} T / m_{\rm p}}$ the Doppler
broadening parameter, $m_{\rm p} = 2 m_{\rm H}$ the mass of the molecule
species, and $\alpha=1.1$. We compute the factor $f_{\rm
  shield,H_2,HD}$ that describes shielding of HD by $\rm H_2$ using
(\citealp{Wolcott2011b})
\begin{equation}
f_{\rm shield,H_2,HD} = \frac{1}{(1+x)^{0.238}} \exp{(-5.2 \times 10^{-3} x)},
\end{equation}
where $x \equiv N_{\rm H_2}/2.34 \times 10^{19} $ $\rm cm^{-2}$.  The
column density $N$ is estimated using $N = n L_{\rm char}$, where $n$
is the number density of the molecule species and $L_{\rm char}$ the
characteristic length scale. We set the latter equal to the local
Jeans length, an approximation that is appropriate for self-gravitating
systems (\citealp{Schaye2001b}; \citealp{Schaye2001a}) and that 
performs well in comparison with more sophisticated approaches 
(e.g., \citealp{Wolcott2011b}).
\par
For reference, the LW intensity of a single Pop~III star at
distance $r$, computed in the optically thin limit, is $J_{\rm LW, 21}
\approx 0.1 (r/1.6\kpc)^{-2}$, where $J_{\rm LW, 21} \equiv J_{\rm LW} /
(10^{-21} \rm erg\ s^{-1} cm^{-2} Hz^{-1} sr^{-1})$. The LW
flux from an accreting BH depends on the accretion rate and the BH
mass. For a $100\msun$ BH with typical Bondi-Hoyle accretion rates of $10^{-8} \Msunyri$
we have $J_{\rm LW, 21} \approx 5.3\times 10^{-5} (r/1.6\kpc)^{-2}$. The
flux of LW radiation implied by a single HMXB with a BH mass of $100 \Msun$ and 
accreting at the Eddington rate is $J_{\rm LW, 21} \approx 4\times 10^{-3}
(r/1.6\kpc)^{-2}$.

\section{Results}
\label{Sec:Results}
In the following, we investigate how the impact of radiative feedback
on Pop~III star formation and the early IGM
varies between the simulations with and without HMXBs. 
First, in
Section~\ref{Sec:FC}, we discuss the feedback of the first HMXB on the
gas in and around its minihalo host. Then, in
Section~\ref{Sec:SFH}, we investigate how the presence of HMXBs affects
the cosmic star formation history. In Section~\ref{Sec:IGM}, we 
study the effects on the IGM, and in Section~\ref{Sec:Halos}, we 
study the effects on the gas in haloes in the two simulations with
and without HMXBs. In Sections~\ref{Sec:Reion} and
\ref{Sec:BHgrowth} we discuss the implications of feedback
from HMXBs for reionization and BH growth.
\par
We emphasize that both simulations account for the radiative
feedback from Pop~III stars and from BHs accreting diffuse halo
gas. The comparison presented here thus elucidates the impact of
radiative feedback by HMXBs complementing the radiative feedback from
Pop~III stars and miniquasars in a simulation of the formation of
the first stars and galaxies.

\subsection{Gas properties in the first minihalo}
\label{Sec:FC}

\begin{figure}
    \includegraphics[width=75mm]{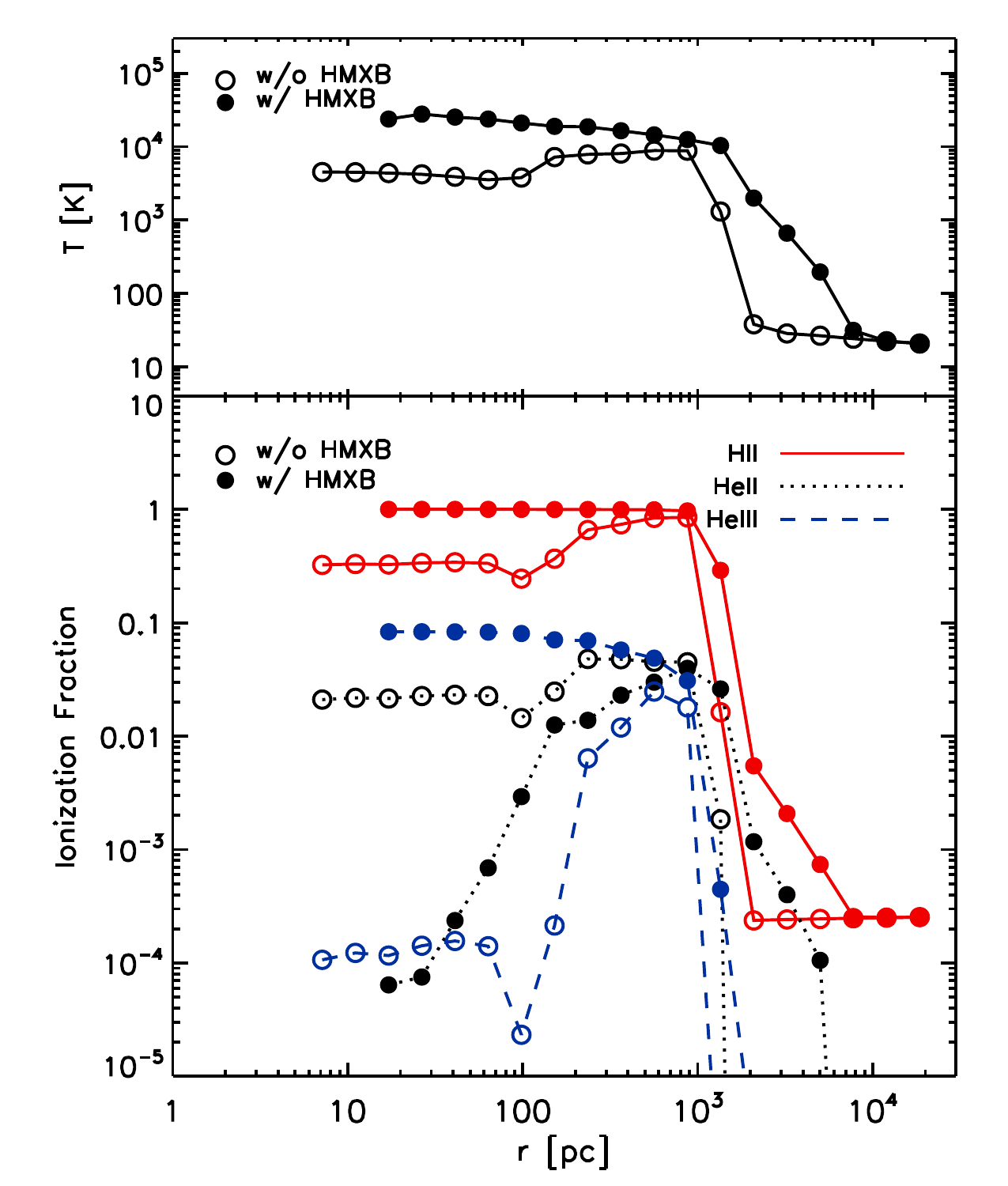}
    \caption{Spherically averaged profiles of the temperature and
      ionization fractions in the simulations with (open circles) and
      without (filled circles) HMXBs at $2 \Myr$ after the formation
      of the BH, the same time as in Figure~\ref{Fig:FirstStarSlices}. In the simulation without
      HMXBs, the relic HII region left behind the first Pop~III star
      has started to recombine and cool. Recombination is enhanced in
      the dense photo-evaporation shock located at $r = 100 \pc$. The HMXB keeps the gas inside the relic stellar HII region
      ionized and hot, and preionizes and preheats the IGM ahead of the
      stellar ionization front. Ionization by the hard HMXB spectrum implies a
      strong increase in the fraction of doubly ionized helium. \label{Fig:ifront}}
\end{figure}

In both runs, the first stellar binary forms at $z=28$; the runs are identical 
up to this redshift. The formation time and the size of the HII region with 
radius $\sim 2-3 \kpc$ are consistent with previous works
\citep[e.g.][]{Kitayama2004,Yoshida2007a, Abel2007,Greif2009}.  The
density profile of the surrounding gas and specifically the location of 
the shock driven by the photoevaporating gas from the minihalo host 
are in good agreement with the results of our earlier simulations in
\citet{Jeon2012} which started from identical initial conditions.
\par
In Figure~\ref{Fig:FirstStarSlices}, we compare images of the gas
temperature 2 Myr after the collapse of the primary into a BH of mass 
$100 \Msun$ in the two simulations. In both simulations, the gas in the relic stellar HII
region has been heated to $\sim10^4 \K$ by the Pop~III progenitor. 
In the simulation without HMXBs, accretion of gas on the BH 
occurs at a rate too low to sustain the high ionization of the gas (see Figure~\ref{Fig:acc} 
discussed in Section~\ref{Sec:BHgrowth}), and the relic HII region 
recombines and cools. In the simulation with HMXBs, on the
other hand, thanks to the large X-ray luminosity of the first HMXB,
the gas in the relic HII region is kept highly ionized and at $\sim 10^4 \K$, and in addition, there is a
significant heating of the gas ahead of the stellar ionization front.
\par
In Figure~\ref{Fig:ifront} we quantify the impact of the first HMXB on
the spherically averaged properties of the halo gas at $2 \Myr$ after
the collapse of the primary into a BH, the same time as in 
Figure~\ref{Fig:FirstStarSlices}. Owing to the slight 
overdensity at the position of the photoevaporation shock, gas
recombination occurs at an increased rate around $\sim 100$ pc from the
source, which explains the dip in the ionized fractions. The enhanced
ionization beyond the I-front in the presence of
the HMXB is mostly due to secondary ionization of HI by electrons
generated in absorptions of X-rays by HeI (Figure~\ref{Fig:secion}; \citealp{Machacek2003}; \citealp{Madau2004}). 
The efficient ionization by X-rays emitted by the HMXB
implies a strong increase in the abundances of HeIII with respect to
that in the simulation without HMXBs. The radii of the HeII and HeIII
regions are only slightly smaller than that of the HII region,
consistent with the results in \citet{Venkatesan2011}.
\par

\subsection{Star formation history}
\label{Sec:SFH}

\par

\begin{figure}
   \includegraphics[width=85mm]{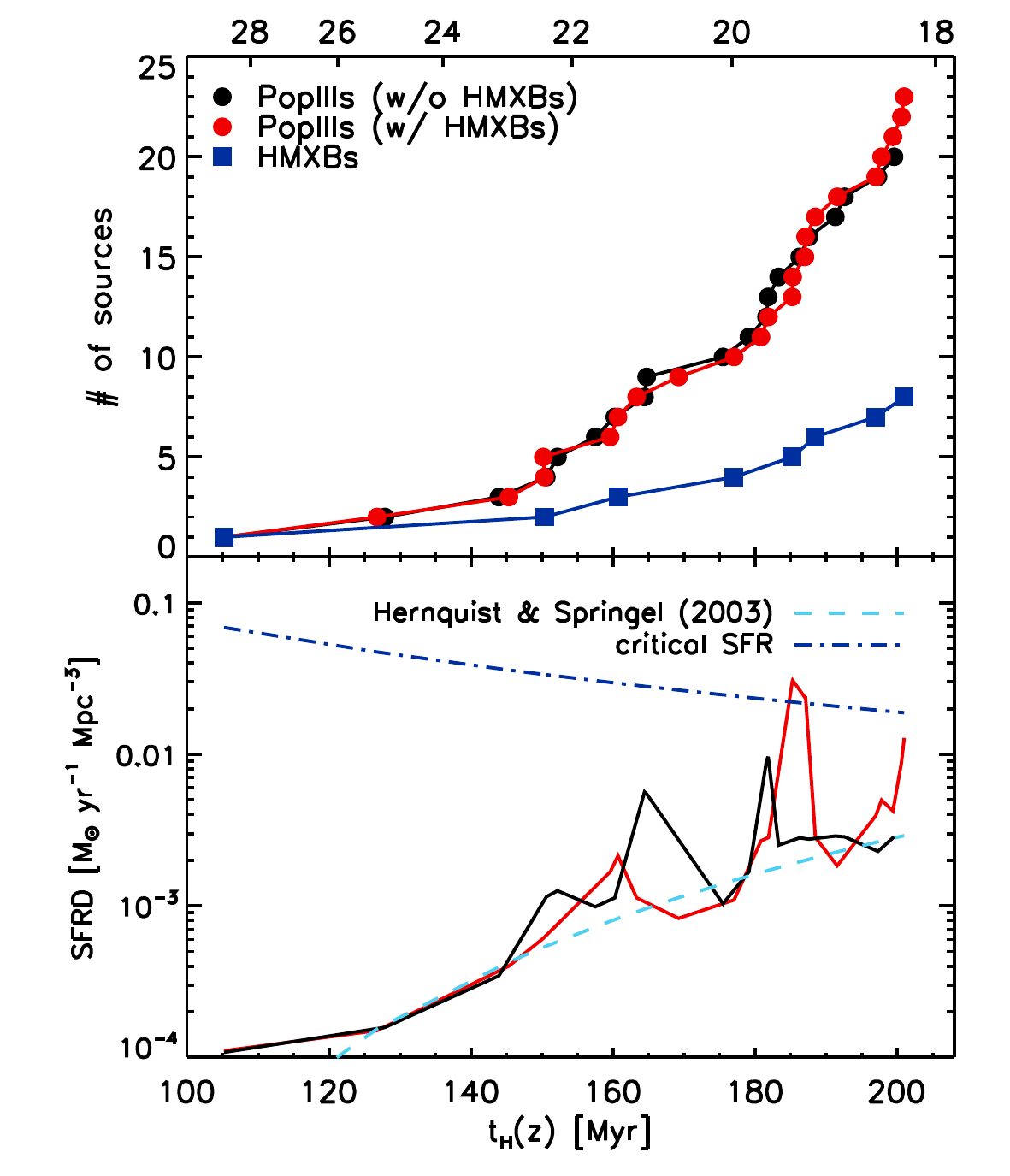}
    \caption{{\it Top:} cumulative number of Pop~III binaries in the
      runs with (red) and without (black) HMXBs, and cumulative number
      of HMXBs (blue squares). The number of stars formed is similar
      in the two simulations, and hence there is no strong net
      feedback by HMXBs. {\it Bottom:} comoving star formation rate
      densities in the two simulations. For comparison, we also show
      the fit to the star formation rate density by \citet[][light
        blue dashed]{Hernquist2003}. The blue dash-dotted curve shows
      a lower limit to the critical star formation rate density needed
      to sustain ionization of the IGM, computed using Equation~\ref{Eq:sfrcrit}
      and assuming $C/f_{\rm esc} = 1$.  \label{Fig:sfr}}
\end{figure} 

\begin{figure*}

    \includegraphics[width=110mm]{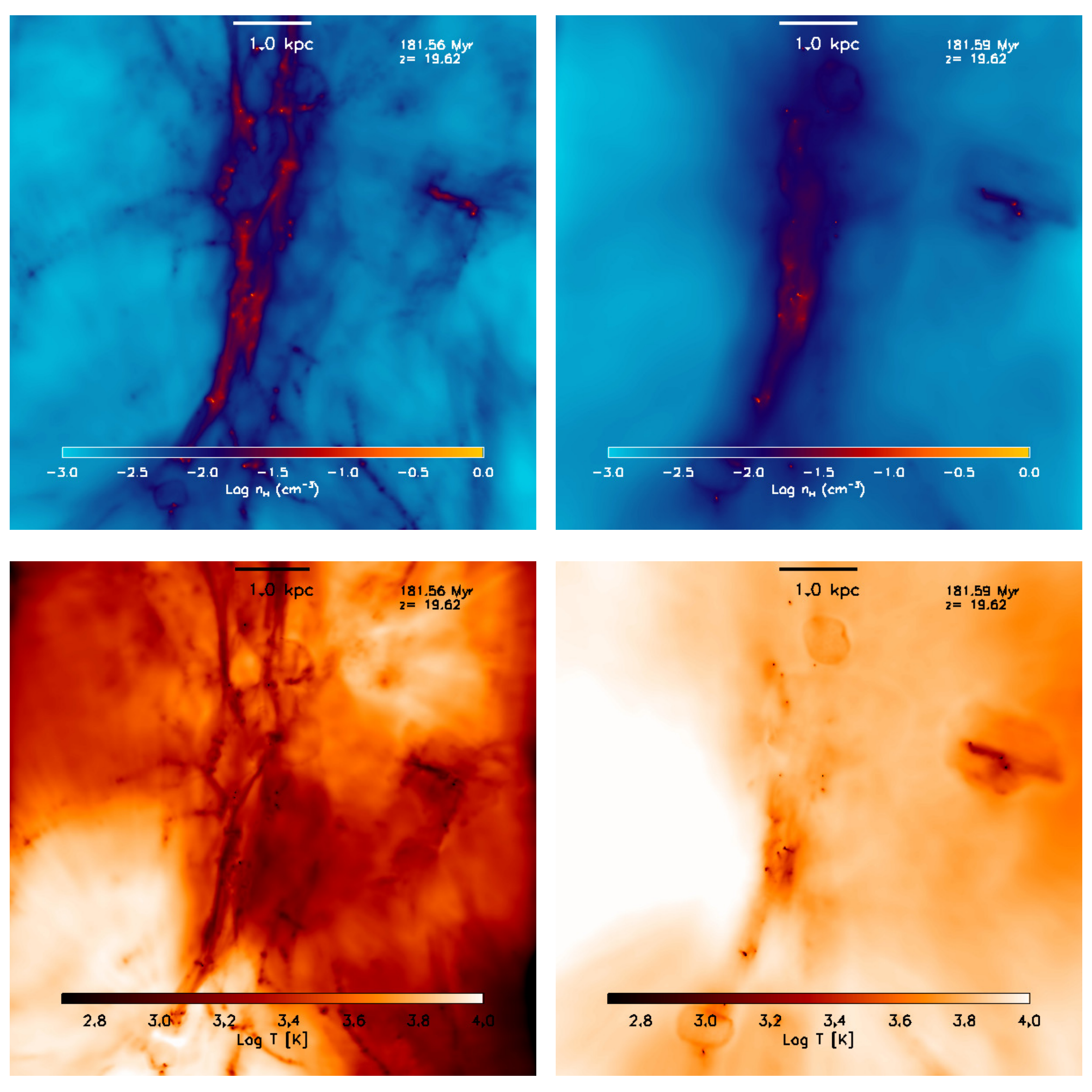}
    \caption{Hydrogen number densities ({\it top}) and temperatures
      ({\it bottom}) at $z = 19.6$, averaged along the line of sight,
      in a cubical cutout of linear extent $300 \ckpc$ centred on the
      first sink particle, in the simulations without ({\it left}) and
      with ({\it right}) HMXBs. The stronger photo-ionization heating in the
      presence of HMXBs and the associated increase in the thermal
      gas pressure imply a stronger smoothing of the gas density
      field.
\label{Fig:temp}}
\end{figure*}

It has been suggested that preionization of gas by the absorption of 
X-rays may promote the formation of
molecular hydrogen and increase the ability of the gas to cool and
form stars (\citealp{Haiman2000}; see also, e.g., \citealp{Ricotti2001}). Several studies
have now confirmed this possibility (e.g., \citealp{Machacek2003};
\citealp{Glover2003}; \citealp{Kuhlen2005}). However, the positive feedback from X-ray
ionization on the formation of molecular hydrogen needs to compete with the negative radiative
feedback caused by pressure-smoothing in the photoheated gas (e.g., \citealp{Shapiro1994}; 
\citealp{Thoul1996}; \citealp{Gnedin2000}; \citealp{Dijkstra2004a}; \citealp{Mesinger2008}; \citealp{Okamoto2008}; \citealp{Wise2008}; 
\citealp{PS2009}; \citealp{Petkova2011}; \citealp{Sobacchi2013}), 
and also with the negative feedback from LW photons
dissociating molecular hydrogen (e.g., \citealp{Haiman1997}; \citealp{Machacek2001}; 
\citealp{Ahn2007}; \citealp{Oshea2008}). Accordingly, most
investigations found at most a modest net positive impact of X-ray ionization on gas
collapse, even in the absence of a strong LW background
(e.g., \citealp{Glover2003}; \citealp{Kuhlen2005}). 
Consistent with these earlier works, in \citet{Jeon2012}, we found that 
X-ray ionization by the first BHs generates at most a modest increase in
the star formation rate in simulations of high-redshift minihaloes. The
exception was an initial brief phase of significantly increased star
formation in an idealized simulation of ionization by a HMXB.
\par 
Figure~\ref{Fig:sfr} compares the formation histories of Pop~III stars
and the associated comoving star formation rate densities in our
current simulations with and without HMXBs. There is little difference
between the two simulations, and hence we find no significant net
radiative feedback from HMXBs on the star formation rate. This is consistent with the lack of a
strong net radiative feedback on the star formation history from
either accreting BHs or HMXBs at $z \lesssim 20$ in
 \citet{Jeon2012}. However, our current simulations do
not reproduce the strong early enhancement in the star formation rate
at $z \gtrsim 20$ in the presence of HMXBs seen in the simulation BHB 
of \citet{Jeon2012}, despite their identical initial
conditions. The lack of an early phase of a net positive feedback 
by HMXBs on star formation likely owes to our adoption 
of a physically more realistic model of the formation of HMXBs 
in the current simulations, as we now explain.
\par
In \citet{Jeon2012}, we followed the radiative feedback from a single
HMXB continuously emitting X-ray photons throughout the simulation. In
the current work, on the other hand, there are 4 HMXBs forming at $z >
20$ (see Figure~\ref{Fig:sfr}). However, each of them emits radiation
for only a brief interval of $2 \Myr$, which in total corresponds to
just about a tenth of the time during which the HMXB was active in our
previous simulation at $z > 20$.  The lack of a net positive feedback
on early star formation in the current simulation is consistent with
this reduced duty cycle for emission of X-rays by HMXBs. Indeed,
Figure~\ref{Fig:sfr} shows that near the end of the simulation, at $z
\lesssim 20$, when HMXB formation occurs more frequently at an average
rate of about once in 5 Myr, the star formation rates in the current
simulation with HMXBs are slightly higher than those in the
simulation without HMXBs.
\par

In the current work we find 19 sink particles at the end of the simulation
without HMXBs at $z \approx 18$. This is a larger number
than the 11 sink particles found at the same redshift in our previous
simulation BHS in \citet{Jeon2012} that included persistent X-ray
radiation from an isolated BH miniquasar accreting diffuse halo gas and that started
from identical initial conditions. We identify two key differences in
the simulation methodology to explain this difference in star
formation rates. First, in the current work, we account for
shielding of the primordial molecular gas from LW photons that 
was ignored in the previous simulations. LW feedback is therefore less
effective in the present work than in the previous simulations and this promotes 
gas cooling and condensation.
Second, in \citet{Jeon2012}, we overestimated the radii of HII
regions around Pop~III stars by a factor of $\sim 2$ as a result of a
numerical error discussed in \citet{Jeon2012}. Thus, in the current work, the suppression of star
formation by stellar ionizing radiative feedback 
is expected to be weaker than in the preceding paper.
\par
\begin{figure*}
  \includegraphics[width=0.8\textwidth,clip=true]{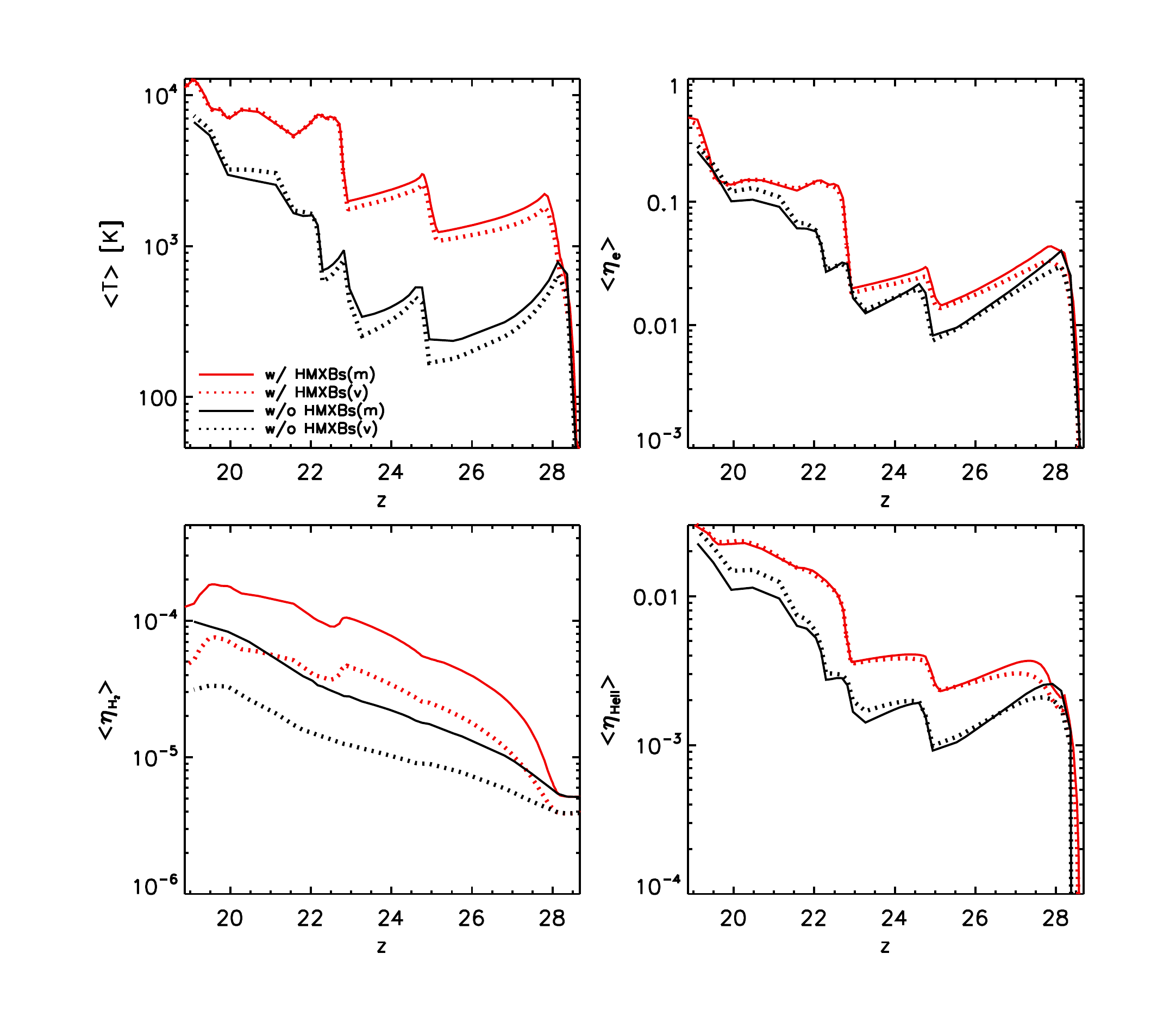}
    \caption{The properties of the gas in the high-resolution region
      of the simulations with (red) and without (black) HMXBs. The
      individual panels show, from left to right and top to bottom,
      temperature, electron fraction, molecular hydrogen fraction, and
      HeII fraction.  Solid (dotted) curves show mass-weighted
      (volume-weighted) averages. HMXBs preionize and preheat the
      gas. The increased electron fraction promotes the formation of
      molecular hydrogen. Ionization by the hard HMXB spectrum generates an
      increased abundance of HeII.
    \label{Fig:avg}}
\end{figure*}
It is interesting to compare the simulated star formation rate
densities with the comoving critical star formation rate density to
sustain ionization against recombinations (\citealp{Madau1999}),
\begin{equation}
\rho_{\rm SFR, crit} = 0.02 \Msun \invyr \Mpc^{-3} \left (\frac{C}{f_{\rm esc}}\right) \left(\frac{1 + z}{20}\right)^3,
\label{Eq:sfrcrit}
\end{equation}
where $C \equiv \langle n_{\rm H}^2 \rangle / \bar{n}_{\rm H}^2$ is
the clumping factor that parametrizes the recombination rate in the
IGM (e.g., \citealp{Madau1999}; \citealp{Miralda2000}; \citealp{Wise2005}; 
\citealp{Pawlik2009}; \citealp{Finlator2012}; \citealp{Shull2012}), the brackets 
indicate the volume-weighted average, $\bar{n}_{\rm H}$ is the cosmological mean gas density, and
$f_{\rm esc}$ is the fraction of ionizing photons that escape haloes
to ionize the IGM (e.g., \citealp{Razoumov2006}; \citealp{Gnedin2008}; 
\citealp{Wise2009}; \citealp{Paardekooper2011}; \citealp{Yajima2011}). Figure~\ref{Fig:sfr} shows that the simulated star
formation rate densities are generally below the critical star
formation rate density, indicating that the reionization of the 
simulated high-resolution region remains incomplete. Because, 
typically, $C > 1$ (see Section~\ref{Sec:IGM}) and, by definition, $f_{\rm esc} \le 1$, 
we have conservatively underestimated the ratios of critical and 
simulated star formation rate densities.

\par 
\par

\subsection{Effects of HMXBs on IGM properties}
\label{Sec:IGM}

Figure~\ref{Fig:temp} shows images of the gas density and temperature
in the high-resolution region centred on the site of formation of the
first stellar binary near the end of the simulation at $z = 19.6$.  In
the simulation that included ionization by HMXBs, the gas is
photoheated to temperatures in excess of a few times $\sim 10^3 \K$
throughout the simulation volume. In the simulation without HMXBs,
which misses the bright HMXB phases of the accreting BHs, on the other
hand, the average temperature is significantly lower, and heating to
comparably high temperatures is limited to volumes inside stellar HII
regions. The higher temperatures in the simulation with HMXBs imply a
cosmological Jeans mass (e.g., \citealp{Shapiro1994}; \citealp{Gnedin2000}; 
\citealp{Wise2008a}) and hence a smoothing of the gas density field
by thermal pressure that is visibly larger than in the simulation
without HMXBs (see Section~\ref{Sec:Reion} for a quantitative discussion).
\par
Figure~\ref{Fig:avg} quantifies the evolution of the properties of the
gas averaged within the high-resolution region in the two
simulations. The additional ionization by the hard radiation emitted 
by HMXBs increases the fraction of HeII with
respect to the fractions found in the absence of HMXBs. The electron
fraction is increased in the simulation with HMXBs over that in the
simulation without HMXBs, largely owing to secondary ionization of
hydrogen (see Section~\ref{Sec:FC}). The fraction of the primary
electron energy that is used in secondary ionizations decreases with
increasing electron fraction and instead heats the gas
(Figure~\ref{Fig:secion}; e.g., \citealp{Shull1985};
\citealp{Furlanetto2010}).  Thus, once the local electron fraction
becomes $\gtrsim 0.1$, further ionization is dominated by
stellar radiation (e.g. \citealp{Mesinger2013}). The increased electron fraction in the simulation with
HMXBs promotes the formation of $\rm H_2$, in agreement with earlier work
(\citealp{Kuhlen2005}; see also, e.g., \citealp{Haiman2000}; \citealp{Machacek2003}; 
\citealp{Ricotti2004}). Near the end of the simulations, when stellar radiation contributes
significantly to photoionization  and dissociation of $\rm
H_2$, the temperatures and species fractions become similar in both simulations.
\par

\subsection{Effect of HMXBs on gas properties in haloes}
\label{Sec:Halos}
\begin{figure*}
   \includegraphics[width=170mm]{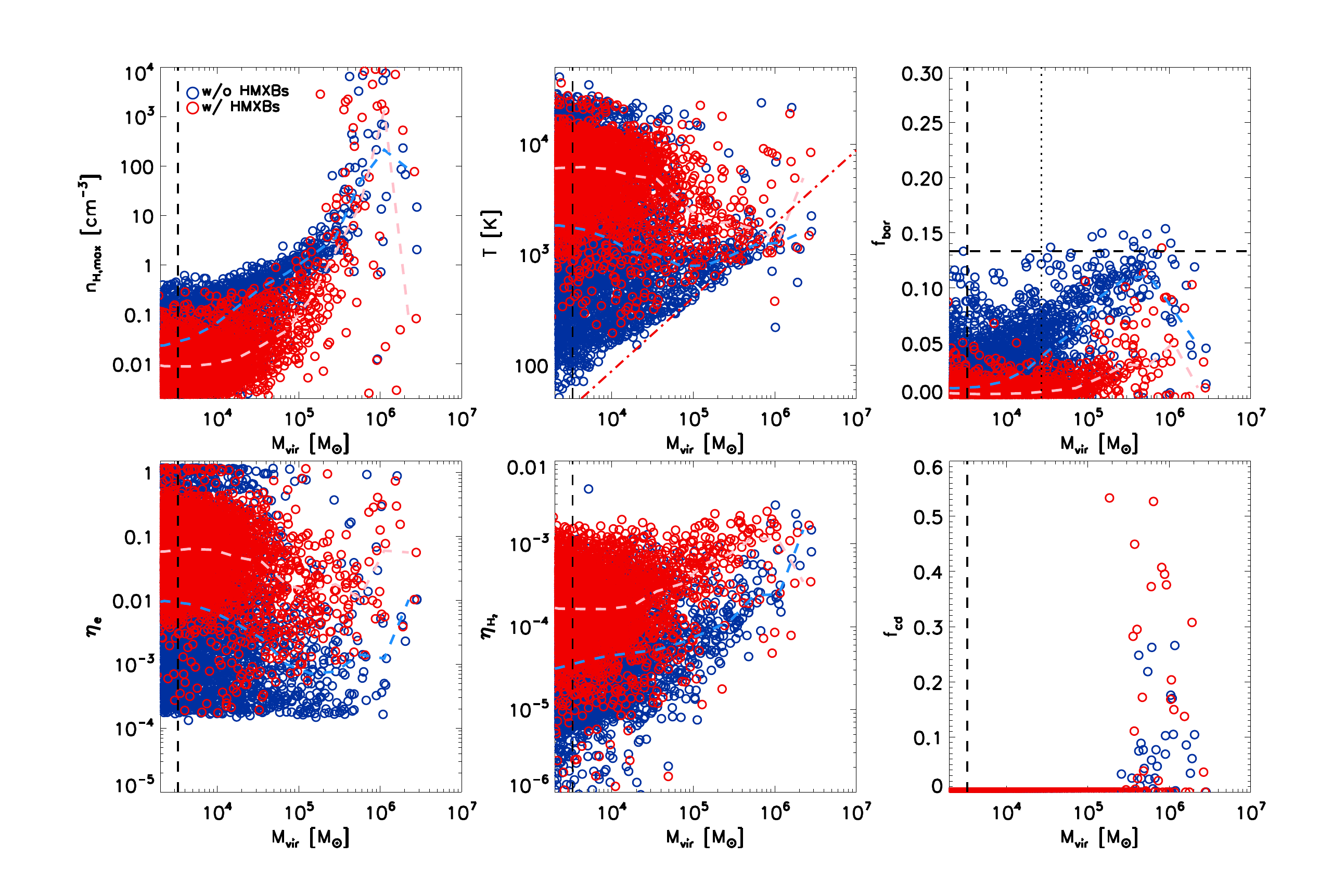}
    \caption{Properties of the gas inside haloes at $z=19.6$ for the
      simulations without (blue circles) and with (red circles) HMXBs.
      The dashed curves of matching color show the median relations to
      guide the eye.  The individual panels show, from left to right
      and top to bottom, the maximum hydrogen number density, the
      mass-weighted average temperature, the baryon fraction, the
      mass-weighted average electron fraction, the mass-weighted
      average molecular hydrogen fraction, and the fraction of cold
      dense gas ($T \le 0.5 T_{\rm vir}$ and $n_{\rm H} \ge
      10^4\bar{n}_{\rm H} \approx 20 \cmci$). The vertical dashed
      lines mark the mass of 100 DM particles. The red dashed-dotted
      line in the top middle panel indicates the virial temperature
      $T_{\rm vir}$ assuming a mean particle weight $\mu = 1.2$, appropriate
      for mostly neutral gas.  The vertical dotted line in the top
      right panel marks the cosmological Jeans mass above which
      gravitational attraction dominates pressure repulsion and haloes
      can accrete gas. In the presence of HMXBs, the temperatures,
      electron fractions, and molecular hydrogen fractions are
      strongly increased. The HMXBs exert a strong feedback on the
      baryon fractions, which are strongly suppressed below the cosmic
      mean (horizontal dashed line) across the range of halo
      masses. However, the feedback from HMXBs on the fraction of cold
      dense gas is only mild, and the number of haloes above the
      critical mass for the onset of cooling and star formation is
      similar with and without HMXBs, consistent with the lack of a
      strong net feedback from HMXBs on the star formation history
      (Figure~\ref{Fig:sfr}).
           \label{Fig:haloes}}
\end{figure*}

In Figure~\ref{Fig:haloes} we compare the properties of the gas inside
haloes in the simulations with and without HMXBs. Halos were extracted
from the high-resolution region displayed in Figure~\ref{Fig:temp} at
$z=19.6$ using the halo finder SUBFIND (\citealp{Springel2001}) and
adopting a linking length equal to a fifth of the mean inter-particle
distance in the high-resolution region to generate friends-of-friends
(FOF) particle group catalogues. Each circle in Figure~\ref{Fig:haloes}
shows the mass-weighted average of the corresponding gas particle
property inside the virial sphere with radius $r_{200}$ centred on
the most-bound particle of the main subhalo inside a given FOF
particle group.
\par 
Several halo properties are similar in both simulations. For
instance, the maximum hydrogen number density $n_{\rm H, max}$ shows
an abrupt increase with increasing halo mass above $M_{\rm vir}
\gtrsim 3\times 10^5 \Msun$. This is caused by the increased ability
of haloes above this mass to form $\rm H_2$ and to cool and condense the
gas (e.g., \citealp{Tegmark1997}; \citealp{Machacek2001}; 
\citealp{Yoshida2003}; \citealp{Oshea2008}). Indeed, the characteristic fraction of
molecular hydrogen features an increase with halo mass near this mass
scale. The lowest mass haloes have characteristic temperatures in
excess of the virial temperature expected for their mass as their gas
is photoionized by stellar radiation from nearby stars. Halos with
mass below the cosmological Jeans mass, evaluated at the cosmic mean
density and the cosmic mean temperature expected in the absence of
star formation  (e.g., \citealp{Gnedin2000}), have baryon fractions 
substantially below the cosmic mean. The most massive minihaloes achieve baryon fractions closer to
the cosmic mean, but also at these masses the scatter is large, owing
to stellar radiative feedback and gravitational and gas-dynamical
interactions between haloes (e.g., \citealp{Wise2008}; \citealp{Wise2012}; \citealp{Pawlik2013}).
\par
The presence of HMXBs leaves clear signatures in the halo gas
properties. The characteristic temperatures and electron fractions are
significantly increased at all mass scales. The increased electron
fraction results in an enhancement of the molecular hydrogen fraction
by an order of magnitude compared to that in the simulation without
HMXBs in all but the most massive haloes. Furthermore, the baryon
fractions are significantly lower in the simulation with HMXBs than
without, even at the largest halo masses. The reason for this is that
X-ray preheating helps stellar radiative feedback by evaporating
halo gas and suppressing baryonic infall onto haloes (e.g., \citealp{Machacek2003};
\citealp{Kuhlen2005}; \citealp{Jeon2012}). However, the presence of
HMXBs does not hinder the build-up of cold dense gas cores in the
centres of haloes that fuel the formation of stars. Here, the cold and
dense gas is defined as the gas with $T\lesssim 0.5 T_{\rm vir}$ and
$n_{\rm H} \gtrsim 10^4 \bar{n}_{\rm H}$ where $\bar{n}_{\rm H}
\approx 1.7 \times 10^{-3} \cmci$ is the cosmic mean gas density at $z
= 19.6$ (cf. \citealp{Machacek2003}).
\par
The mild impact on the cold dense gas suggests that the reduction in
the baryon fraction originates mostly in the outer lower density
regions of the haloes and is consistent with the lack of a strong net
effect of HMXBs on the star formation history. In principle, the
increase in $\rm H_2$ can help the collapse of the gas in haloes by
increasing the ability of the primordial gas to cool. 
Indeed, the simulation including HMXBs shows an increased
amount of cold dense gas for haloes above the critical mass for the
onset of efficient cooling. However, the number of haloes with cold gas
fraction $\gtrsim 0.1$ is similar in both our simulations. Moreover, 
the critical mass for the onset of cooling is insensitive
to the presence of HMXBs, and above this mass molecular hydrogen
formation is already very efficient in the absence of additional
positive feedback. 
\par
Our results are consistent with the results of \citet{Machacek2003}
and \citet{Kuhlen2005}, both of which investigated the feedback from X-ray
ionization on structure formation and halo gas properties in
cosmological simulations of high-redshift minihaloes. Varying the
strength of the X-ray flux by two orders of magnitude with respect to
the intensity of the imposed LW background, \citet{Machacek2003} found
only a mild effect on the ability of gas to cool and collapse to high
densities. Even a high X-ray flux was found to be
insufficient to substantially increase the fraction of cold dense gas
or to significantly lower the critical halo mass for the onset of
efficient gas cooling. A similar insensitivity of the cold dense gas
fractions to X-ray ionization was found in \citet{Kuhlen2005} and
attributed to the competition with the negative feedback from
photoheating.
\par

\subsection{Reionization}
\label{Sec:Reion}
\begin{figure}
  \includegraphics[width=0.5\textwidth,clip=true]{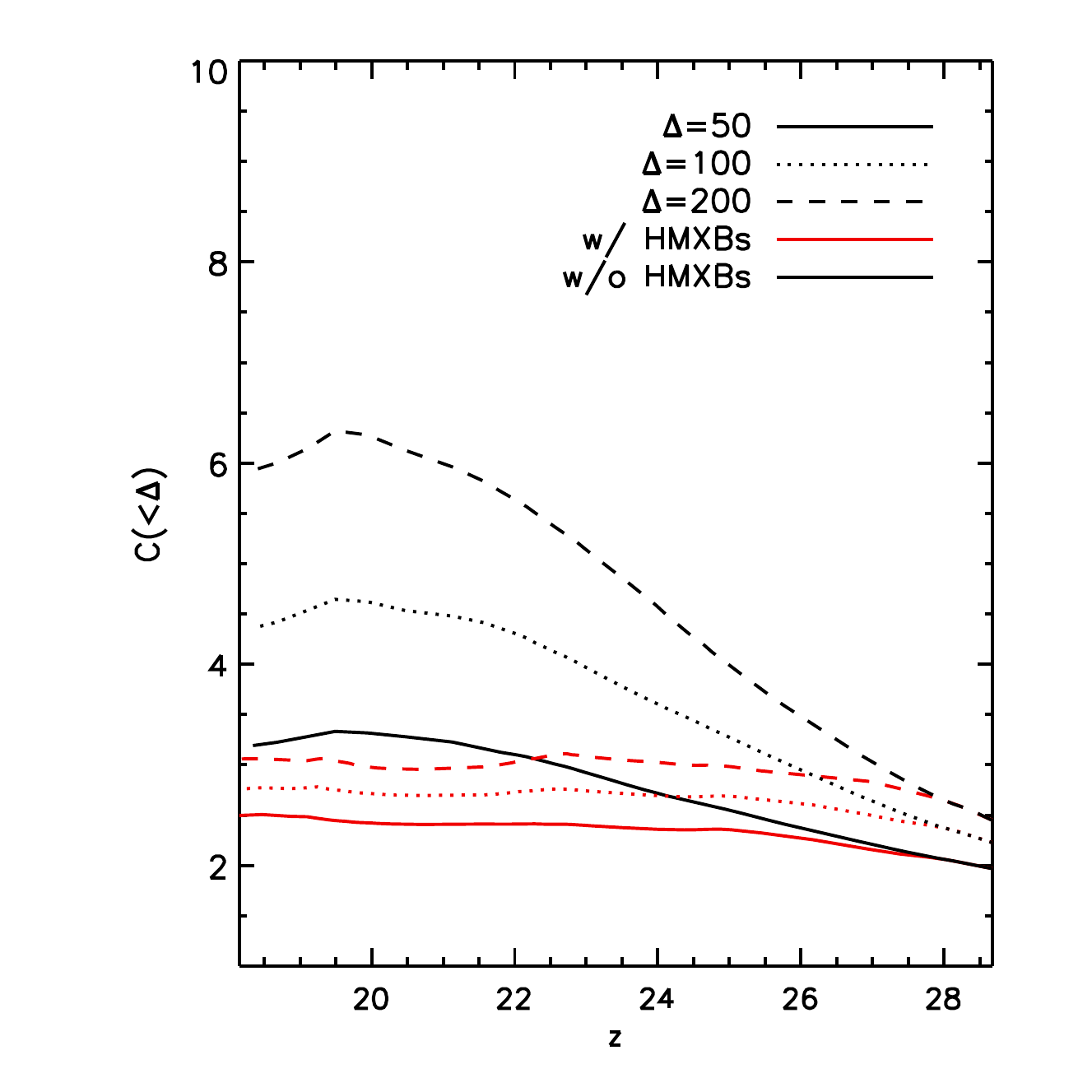}
  \caption{The volume-weighted clumping factor of gas with
      overdensities $\le \Delta$. HMXBs preionize
      and preheat the gas, implying a strong smoothing of the diffuse
      IGM, reducing its clumpiness. This makes it easier to keep the
      gas ionized, and hence provides a positive feedback on reionization.
    \label{Fig:clumping}}
\end{figure}

It is generally accepted that the stellar radiation from galaxies is
mainly responsible for the reionization of the Universe
(for reviews see, e.g., \citealp{Barkana2001}; \citealp{Barkana2009}; 
\citealp{Robertson2010}). However, X-ray sources,
such as BHs, HMXBs, and structure formation and supernova remnant shocks
\citep[e.g.][]{Oh2001, Madau2004, Ricotti2004, Ripamonti2008,
  Power2009, Johnson2011b, McQuinn2012, Mirabel2011, Haiman2011,
  McQuinn2012, Mesinger2013, Fialkov2014}, could have established a floor of
ionized fractions of $\sim 10$ per cent nearly uniformly throughout the
Universe and preheated the IGM to a few times $ \sim 10^3 \K$
\citep[e.g.][]{Oh2001, Volonteri2009, Johnson2011b, McQuinn2012,
  Mesinger2013}. The associated X-ray feedback may delay reionization 
  by impeding galaxy formation due to photoionization heating 
(e.g., \citealp{Ef1992}; \citealp{Shapiro1994}; 
\citealp{Gnedin2000}; \citealp{Finlator2011}). Or it may
accelerate it, by reducing the clumping factor of the IGM, and hence
the number of ionizing photons required to keep the Universe ionized
(e.g., \citealp{Wise2005}; \citealp{Pawlik2009}), or by promoting the formation of 
molecular hydrogen (\citealp{Haiman2000}; see 
also, e.g., \citealp{Ricotti2001}; \citealp{Oh2002})
\par 
In \citet{Jeon2012} we showed that radiative feedback from the Pop~III
progenitor stars strongly reduces the Bondi-Hoyle rates at which BHs
accrete diffuse gas in the first minihaloes (see also, e.g., 
\citealp{Alvarez2009}). Accordingly, the initial
radiative impact of these BHs on the gas was found to be small. 
This is consistent with the results from our
current simulation without HMXBs, in which many properties of the gas
in the IGM and in the haloes are similar to those found in
simulations of the first stars and galaxies that do not include 
the emission of X-rays by accreting BHs. The X-ray luminosities of HMXBs, on the other 
hand, are insensitive to the radiative feedback from the Pop~III progenitor 
stars, as mass accretion occurs at high, near-Eddington rates directly from the stellar companion. As
a result, we find that HMXBs substantially preionize and preheat the
gas ahead of the ionization fronts driven by the first stars already
at early times at which the radiative impact from BHs accreting
diffuse gas is still negligible.

\begin{figure}
   \includegraphics[width=85mm]{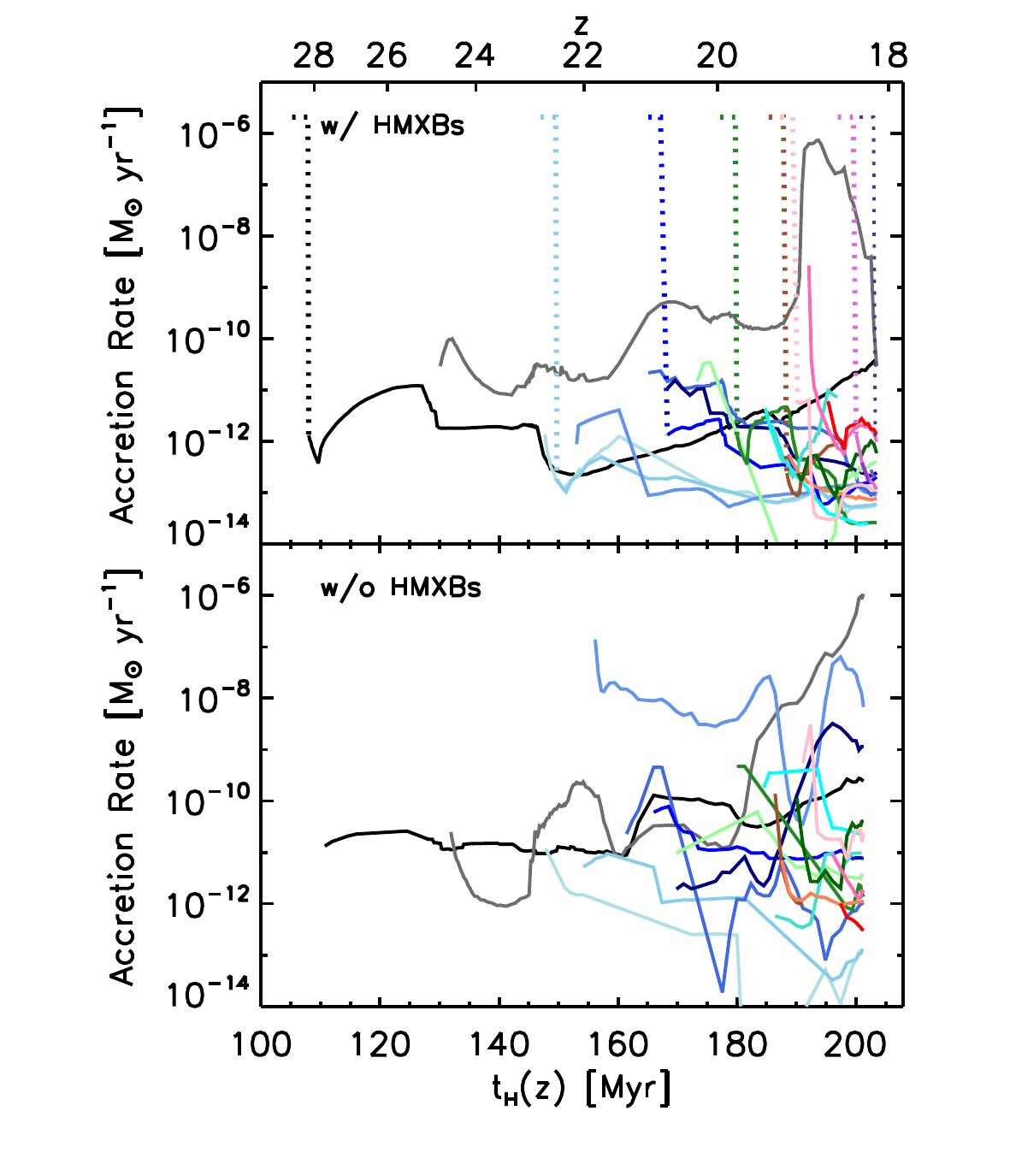}
    \caption{BH accretion rates in the simulations with ({\it top}) and
      without ({\it bottom}) HMXBs. The solid curves show the Bondi-Hoyle rates at which isolated 
      BHs accrete diffuse halo gas. The dotted curves show the Eddington rates, $2.2
      \times 10^{-6} \Msunyri (M_{\rm BH} / 100 \Msun)$, of BHs accreting gas from the stellar companion
      in an HMXB. Stellar radiative feedback from the Pop~III progenitor stars strongly reduces the rate of 
      accretion of diffuse gas several orders of magnitude below the Eddington rate. 
      HMXB feedback typically leads to an additional suppression of 
      the Bondi-Hoyle accretion rates by a factor of a few. \label{Fig:acc}}
\end{figure}

\begin{figure}
   \includegraphics[width=85mm]{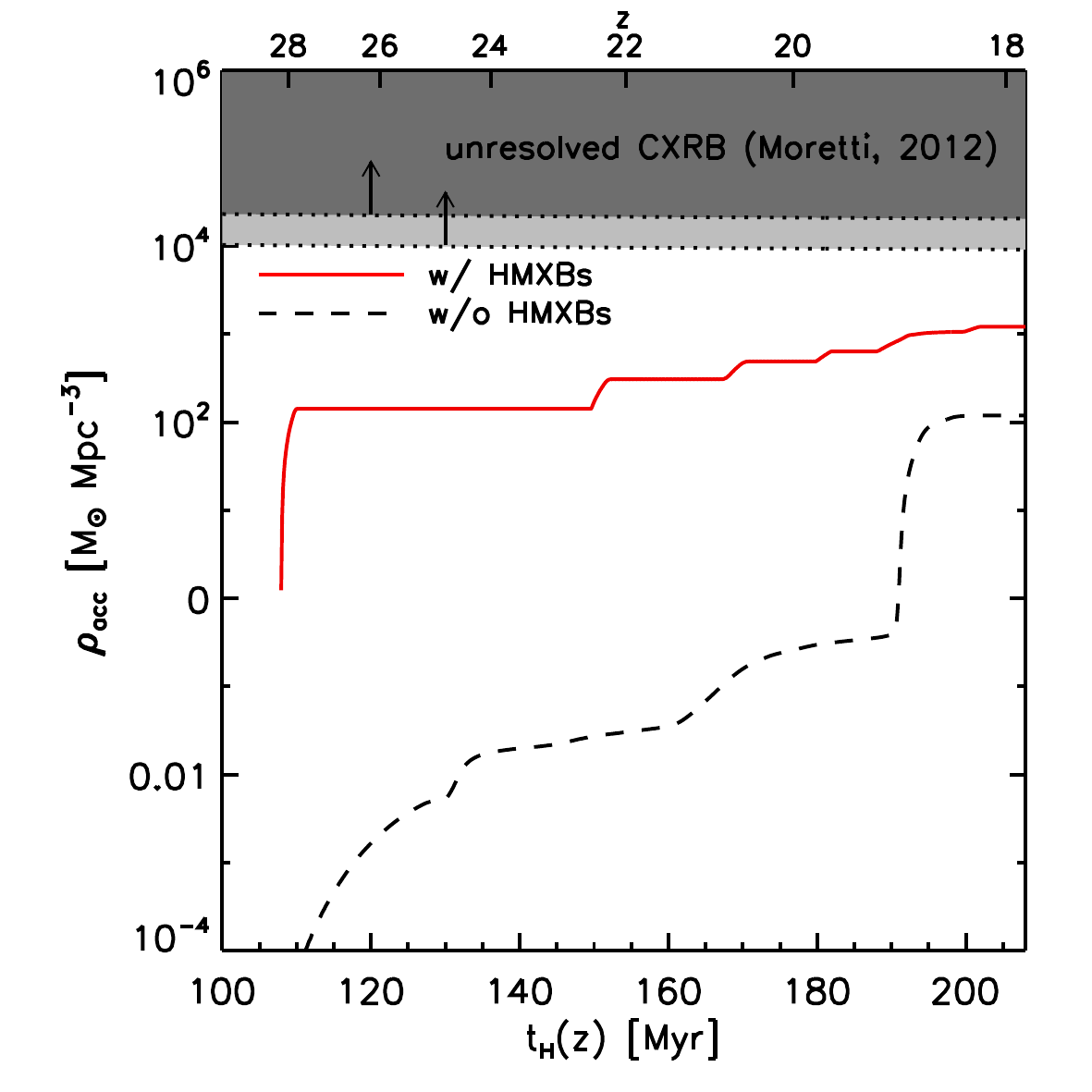}
    \caption{Comoving mass density accreted
      onto BHs in the simulations with (red solid) and without
      (black dashed) HMXBs. The shaded area marks the upper limit on the mass accretion density 
      above $z > 5$ allowed by the unresolved CXRB (\citealp{Moretti2012}), as computed in \citet{Salvaterra2012}, 
      without (dark shaded) and with (light shaded) 
      the contribution from faint sources at $z<5$. The simulated comoving mass 
      accretion densities are consistent with the observational constraints. However, they may  
      violate these constraints if extrapolated to $z = 5$, implying a tension between CXRB limits and models of
      BH growth highlighted in \citet{Salvaterra2012}. 
      \label{Fig:rho}}
\end{figure}

\par
A key quantity for discussing reionization is the clumping factor $C$,
which parametrizes the recombination rate in the IGM (see
Section~\ref{Sec:SFH}).  Figure~\ref{Fig:clumping} shows the clumping
factor computed by considering only gas with overdensities $\Delta
\equiv n_{\rm H}/\bar{n}_{\rm H} \le \Delta_{\rm IGM}$ characteristic
of the IGM, where $\bar{n}_{\rm H}$ is the cosmic mean density. We
adopt $\Delta_{\rm IGM} = 100$ as reference, and show the sensitivity of the 
clumping factor to changes in $\Delta_{\rm IGM}$ by a factor of 2 above and below the 
reference value. The clumping factor is substantially smaller in the run with
HMXBs than without, confirming the visual impression of pressure
smoothing in Figure~\ref{Fig:temp}.  The evolution of the clumping
factor in the absence of HMXBs, when gas temperatures are determined
mostly by stellar radiative feedback, is consistent with that in
\citet{Wise2008a}. In their simulations, radiative feedback from the
first stars decreased the gas clumpiness by $\sim 25$ per cent
compared to that in a comparison run without star formation. Our work
demonstrates that in the presence of X-ray heating by HMXBs, the gas
clumpiness can be reduced by an additional significant factor of $\sim
2$.
\par
In our current simulations, the rate at which stars form is
insensitive to the presence of HMXBs (Figure~\ref{Fig:sfr}). As discussed earlier, this is
likely because the positive feedback by X-ray ionization on the
formation of molecular hydrogen and the cooling of the primordial gas
is partially offset by the negative feedback from photoheating and
Jeans smoothing (Figure~\ref{Fig:haloes}; e.g., \citealp{Machacek2003};
\citealp{Kuhlen2005}). On the other hand, the presence of HMXBs leads
to a significant reduction in the clumping factor of the IGM
(Figure~\ref{Fig:clumping}), and hence in the number of ionizing photons
required to sustain ionization in the HII regions blown by the first
stars. Our simulations therefore demonstrate that ionization by HMXBs
can provide a significant net positive feedback on reionization. Such
a net positive feedback has sometimes been assumed to operate in
semi-analytic models (e.g., \citealp{Madau2004}; \citealp{Mirabel2011}). 
\par

\subsection{Black hole growth}
\label{Sec:BHgrowth}

Observations of supermassive BHs (SMBHs) at $z\gtrsim 6$
(for reviews see, e.g., \citealp{Fan2006}; \citealp{Willott2010}; \citealp{Mortlock2011}) 
have triggered the question of their assembly. Various scenarios 
for SMBH formation have been suggested \citep[for a recent review see][]{Volonteri2012}, including the growth
from a Pop~III stellar-mass seed BH
by gas accretion and mergers with other stellar mass BHs. A strong
obstacle to this formation path is provided by the feedback from the first
stars in the galaxies hosting the BHs, impeding gas accretion and
delaying BH growth for $\sim$ 100 Myr \citep[e.g.,][]{JB2007,Pelupessy2007,
  Alvarez2009, Jeon2012, Johnson2013}, and by the feedback from the
BHs themselves (e.g., \citealp{Milos2009a}; \citealp{Alvarez2009};
\citealp{Park2011}; \citealp{Jeon2012}) However, only a small fraction
of all haloes needs to experience runaway BH growth to reach the observed space density of
SMBHs at $z\gtrsim6$ (e.g., \citealp{DijkstraSMBH2008}; \citealp{Tanaka2012}; \citealp{Johnson2013}).
\par

\begin{figure*}
   \includegraphics[width=155mm]{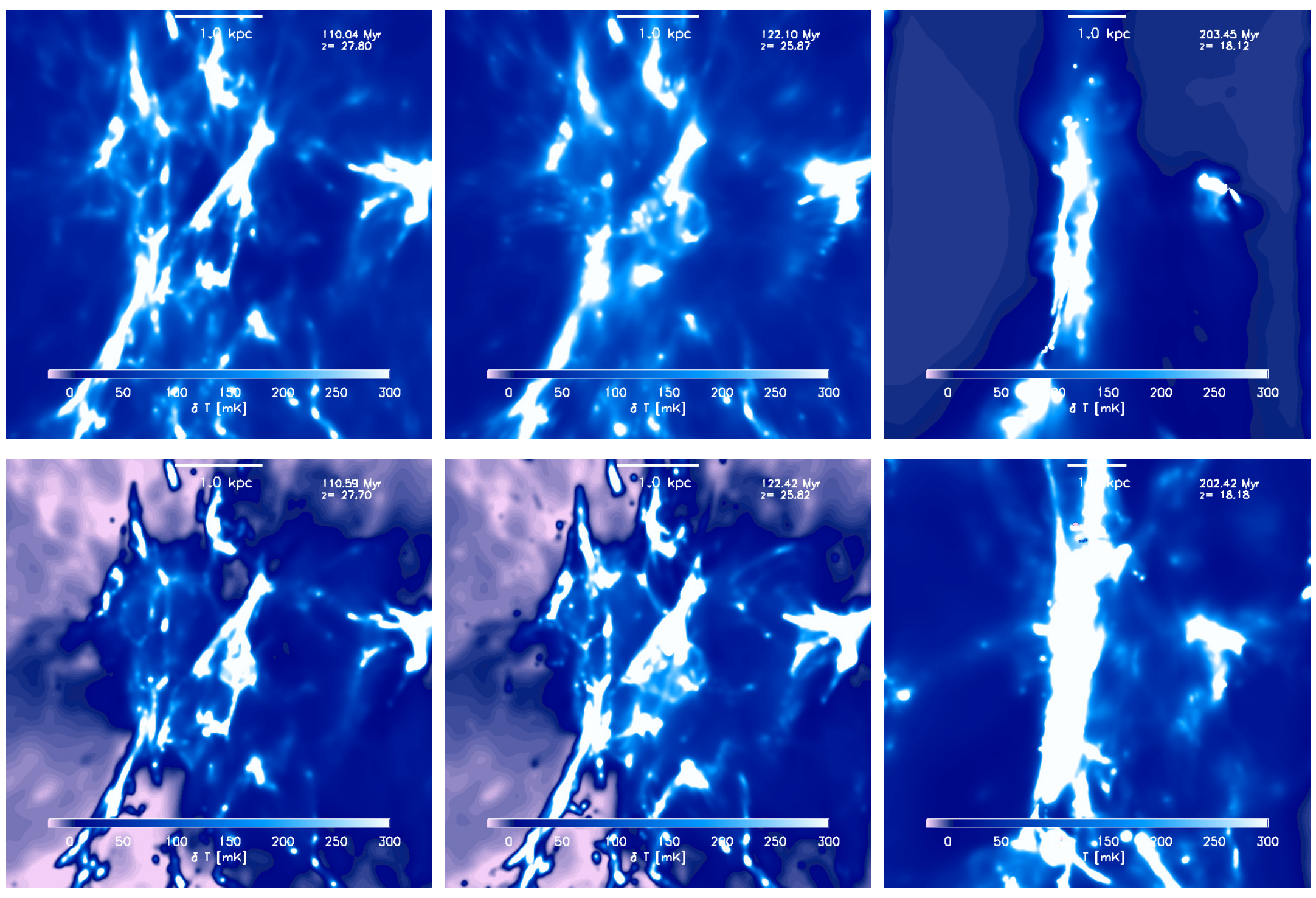}
    \caption{21~cm differential brightness temperatures $\delta T_{\rm
        b}$ at $z=27.8, 25.9$, and $18.1$ (from left to right),
      averaged along the line of sight, in cubical cutouts of linear
      extent $300 \ckpc$ centred on the first sink particle, in the
      simulations with ({\it top}) and without ({\it bottom})
      HMXBs. The computation assumed that the spin temperature is
      equal to the gas temperature, $T_{\rm s} = T_{\rm gas}$, as may be
      expected after the establishment of a Ly-$\alpha$ background by
      the first stars. At early times (left and
      middle panels), in the absence of HMXBs, there is no significant
      preheating of the IGM outside the stellar HII regions, and the
      neutral IGM is seen in 21~cm absorption ($\delta T_{\rm b} <
      0$). In contrast, in the presence of HMXBs, X-ray preheating
      quickly establishes a nearly uniform temperature floor, and the
      neutral gas outside stellar HII regions is seen in 21~cm
      emission ($\delta T_{\rm b} > 0$). At late times (right panels),
      the gas in the simulation with HMXBs is highly ionized and the
      21~cm signal of the IGM nearly vanishes. In contrast, in the
      absence of HMXBs, the IGM outside stellar HII regions remains
      significantly neutral and is visible in 21~cm emission. The
      systematic differences in the differential brightness temperature of the IGM
      may potentially be used to constrain the nature of the first
      sources in upcoming 21~cm observations.
      \label{Fig:21cm}}
\end{figure*}

Figure~\ref{Fig:acc} shows the gas accretion rates onto
BHs in our two simulations with and without HMXBs. The Bondi-Hoyle accretion rates
are up to $\sim 8$ orders of magnitude below the Eddington limit of $2.2
\times 10^{-6} \Msunyri (M_{\rm BH} / 100 \Msun)$. This owes primarily
to the feedback from the Pop~III stars, but in the simulation with
HMXBs, the Bondi-Hoyle accretion rates are suppressed, on average, to still lower
values by feedback from X-ray heating.  Interestingly, in both
simulations, near the final simulation redshift, one of the BHs
reaches diffuse accretion rates close to the Eddington rate, triggered by the passage 
through dense clouds of gas. Such near-Eddington accretion could be
characteristic of BHs residing in rare, high-density peaks in which
self-gravity overcomes the feedback effects (e.g., \citealp{Li2011}).
\par 
Measurements of the cosmic X-ray background (CXRB) have been used to
constrain the population of high-redshift X-ray sources \citep[e.g., ][]
{Dijkstra2004, Ricotti2004, Salvaterra2007, Treister2011, McQuinn2012,
  Salvaterra2012, Dijkstra2012}.  \citet{Salvaterra2012} used the \emph{Swift} X-ray telescope
data in the hard X-ray band \citep{Moretti2012} to obtain a limit of
$\rho_{\rm acc}\lesssim 1.4 \times10^4\msun \rm Mpc^{-3}$ on the mass
density accreted by BHs above $z \ge 5$, adopting the AGN spectra proposed by
\citet{Sazonov2004} and a radiative efficiency $\epsilon=0.1$. The dark shaded
area in Figure~\ref{Fig:rho} shows this upper limit extrapolated to
high redshifts using their equation~7. The light-shaded area shows the
corresponding reduced upper limit, discussed in \citet{Salvaterra2012}, of $\rho_{\rm
  acc}\lesssim 6.6 \times10^3\Msun \Mpc^{-3}$ at $z = 5$, now accounting
for the contribution from faint sources at $z \lesssim 5$, as modeled 
in \citet{Gilli2007}.
\par
The simulated comoving accretion mass densities onto BHs and HMXBs shown
in Figure~\ref{Fig:rho} are computed using
\begin{equation}					  
\rho_{\rm acc} = \int_{z=30}^{z=18} dt\ {\sum_{all}{(\dot{\rho}_{\rm acc,
      BH} + \dot{\rho}_{\rm acc, HMXB} )}},
\end{equation}
where $\dot{\rho}_{\rm acc, BH}$ is the time-dependent accretion mass
density onto BHs accreting diffuse halo gas, and $\dot{\rho}_{\rm acc,
  HMXB}$ is the time-independent accretion rate density onto BHs of
mass $100 \Msun$ inside HMXBs accreting gas from the stellar companion
at the Eddington rate, $2.2 \times 10^{-6} \Msunyri (M_{\rm BH} / 100 \Msun)$, for a duration
of $2 \Myr$ each.  The accretion rate density is larger in the
simulation with HMXBs than in the simulation without HMXBs by up
to several orders of magnitude, as expected from the difference in
accretion rates of HMXBs and miniquasars.
\par
At the final simulation redshift, the accreted mass densities are
still significantly below the CXRB upper limits and hence consistent
with observations. However, if BH growth continues down to lower
redshifts as expected, the accretion mass densities would likely
exceed the CXRB upper limits by $z = 5$. This illustrates the tension
between the CXRB limits and models of BH growth that was highlighted
by \citet{Salvaterra2012}. However, our simulations may overestimate
the accretion mass density since a fraction of the Pop~III stars may
explode in pair instability supernovae instead of collapsing into BHs
(e.g., \citealp{Heger2003}; \citealp{Chatzopoulos2012},
\citealp{Yoon2012}), a process that we have
ignored. Moreover, at lower redshifts, stellar radiative, kinetic and
chemical feedback as well as BH feedback may limit the ability of
galaxies to grow BHs, further reducing the accreted mass densities
(e.g., \citealp{Tanaka2012}). The contribution of redshifted radiation
from high-$z$ sources to the observed $2-10 \keV$ band depends on their
X-ray spectra (see \citealp{Salvaterra2012} for a discussion) and will
be reduced if the typical spectrum cuts off at rest-frame energies
lower than $(40-200)/[(1+z)/20] \keV$.

\section{Impact on 21~cm signature}
\label{Sec:21cm}

The 21~cm line of atomic hydrogen is a promising tool to
understand the reionization history of the Universe (for reviews 
see, e.g., \citealp{Furlanetto2006}; \citealp{Pritchard2012}; \citealp{Zaroubi2013}). The 
relevant observable is the differential brightness temperature, i.e.,  
the 21 cm brightness temperature measured against the CMB temperature $T_{\rm CMB}$, and 
which in the optically thin limit is given by (e.g., \citealp{Madau1997}; \citealp{FurlanettoOh2006})
\begin{equation}
\delta T_{\rm b} = 40 \mK\  (1+\delta)\ \eta_{\rm HI} \left( \frac{1+z}{25} \right)^{1/2}  \left(1-\frac{T_{\rm CMB}}{T_{\rm s}}\right),
\end{equation}
where $\delta \equiv (n_{\rm H} - \bar{n}_{\rm H})/\bar{n}_{\rm H}$ is the density contrast, and 
$T_{\rm s}$ is the spin temperature. The spin temperature is the weighted sum of the kinetic gas
temperature $T_{\rm gas}$ and $T_{\rm CMB}$,
\begin{equation}
	T_{\rm s} = \frac{T_{\rm CMB} + (y_{\rm col} + y_{\alpha}) T_{\rm gas}}{1+y_{\rm col}+ y_{\alpha}},
\label{Eq:t21}
\end{equation}
where $y_{\rm col}$ and $y_{\alpha}$ are, respectively, the collisional and Ly$\alpha$ radiative 
coupling efficiencies (e.g., \citealp{Wouthuysen1952}; \citealp{Field1958}), and we have assumed 
$T_{\rm c} = T_{\rm gas}$ for the color temperature $T_{\rm c}$ of the Ly$\alpha$ photons.
\par
Depending on the spin temperature of the hydrogen atom, the 21~cm line
may be seen in emission ($\delta T_{\rm b} > 0$) or absorption
($\delta T_{\rm b} < 0$).  Estimates of the spin temperature are
complicated by uncertainties in the radiative coupling by Ly$\alpha$
photons that may originate from stellar atmospheres or from recombining 
or excited ions, and whose intensity is difficult to compute because it requires 
solving the RT equation for Ly$\alpha$ photons (e.g., \citealp{Baek2009}; \citealp{Yajima2013}). 
Therefore, most works simply assume either full collisional or full radiative coupling given the 
circumstances \citep[e.g.][]{Greif2009,
  Tokutani2009}. E.g., once the first generation of stars turns on,
Ly$\alpha$ coupling is expected to quickly set the spin temperature 
to the gas temperature \citep[e.g.][]{Chen2008, Ciardi2010, Mesinger2013}. Note from
Equation~\ref{Eq:t21} that for $T_{\rm s} \gg T_{\rm CMB}$, the
differential brightness temperature is independent of the spin
temperature (e.g, \citealp{Scott1990}).
\par
Here, we compute the 21~cm signal assuming the two limiting cases of pure collisional
coupling ($y_\alpha = 0$) and perfect coupling to the gas kinetic temperature ($T_{\rm
  s} = T_{\rm gas}$), due to Ly$\alpha$ scattering via the Wouthuysen-Field effect 
\citep{Wouthuysen1952,  Field1958}, but neglecting any accompanying
Ly$\alpha$ heating of the gas. Such a strong radiative 
coupling is expected after the establishment of a Ly$\alpha$ background 
by the first stars (e.g., \citealp{Kuhlen2006}). The efficiency
of collisional coupling is given by $y_{\rm col} = T_{\ast}(C_{\rm H}
+ C_{e} + C_{p})/(A_{\rm 10} T_{\rm gas})$.  Here, $T_{\ast}=h_{\rm P} \nu_{\rm 
21~cm}/k_{\rm B}=0.0681\K$ is the temperature
associated with the 21~cm hyperfine-structure transition, $A_{\rm 10}=2.85
\times 10^{-15} \rm s^{-1}$ is the Einstein A coefficient and $C_{\rm
  H}$, $C_{e}$, and $C_{p}$ are the de-excitation rates of the
triplet state due to collisions with neutral atoms, electrons and protons,
respectively.  The collisions with neutral atoms are determined by
$C_{\rm H} = n_{\rm H} \kappa$, where $\kappa$ is the effective
single-atom rate coefficient from \citet{Zygelman2005}. The $e$-H
collision term is written as $C_{e} = n_{e} \gamma_{e}$,
where $\gamma_{e}$ is given by \citet{Liszt2001} and
\citet{Smith1966}. The rate coefficient for proton-induced de-excitation is
just 3.2 times larger than that for neutral atoms at $T_{\rm gas} > 30$
K \citep{Smith1966}, $\gamma_{p}=3.2\kappa$, and $C_{p}=n_{p} 
\gamma_{p}$. Excitation by protons is typically unimportant,
as it is much weaker than that by electrons at the same temperature 
(e.g., \citealp{Kuhlen2006}).
\par 

\begin{figure}
   \includegraphics[width=87mm]{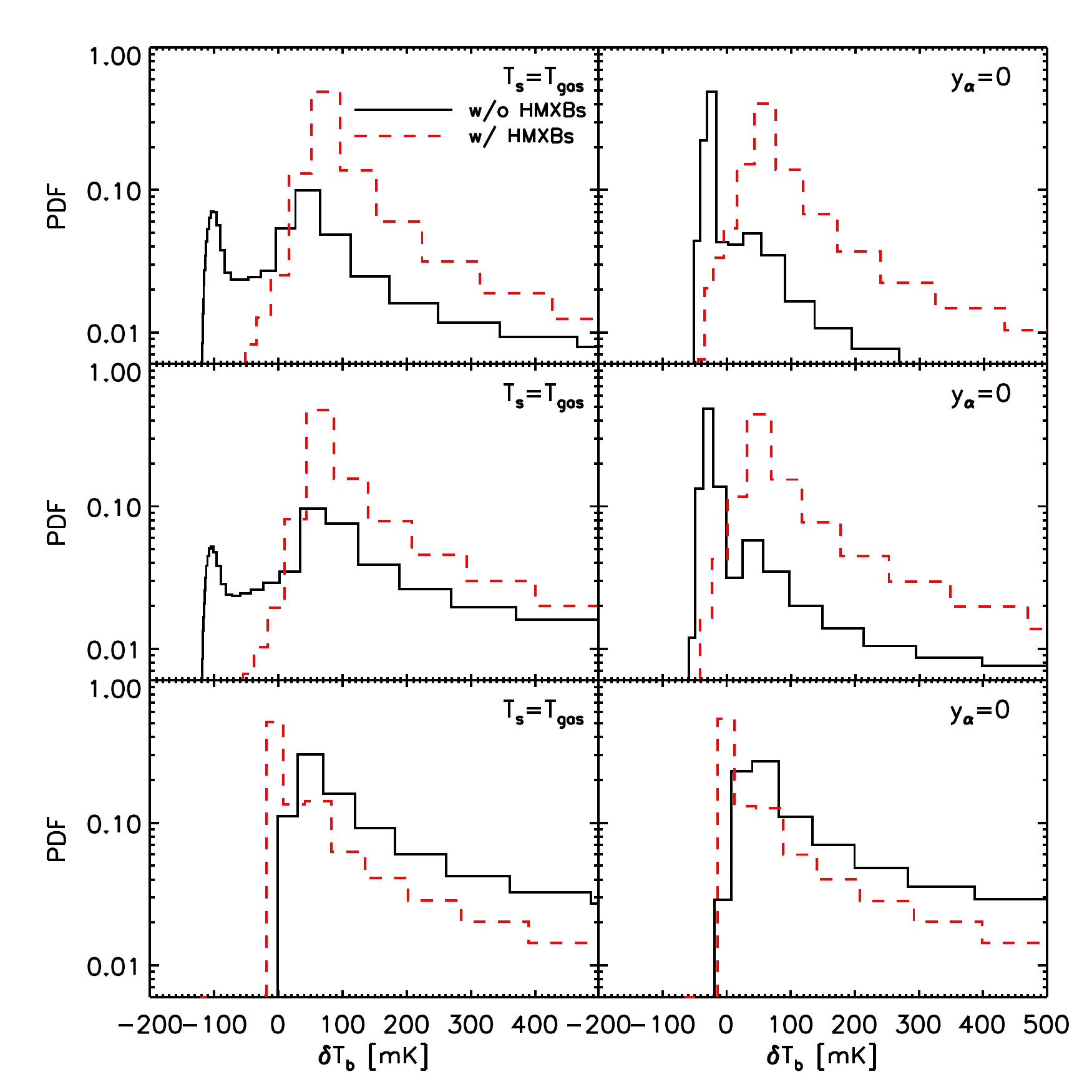}
    \caption{The probability distribution function (PDF) of the
      differential brightness temperature $\delta T_{\rm b}$ at
      $z=27.8, 25.9$, and $18.1$ (from top to bottom), in the
      high-resolution regions of the simulations with (red dashed) and
      without (black solid) HMXBs. Panels in the left column show the
      PDF computed assuming $T_{\rm s} = T_{\rm gas}$, as will be the
      case after the establishment of a  Ly$\alpha$
      background by the first ionizing sources, and panels
      in the right column show the PDF assuming that spin and gas
      temperatures couple only through collisions ($y_{\alpha} = 0$ in Equation~\ref{Eq:t21}).       
   \label{Fig:pdf}}
\end{figure}

We point out that our high-resolution region, corresponding to the angular size of 
$\sim 0.1\arcmin$ and a frequency range of $\sim 0.01 \MHz$ (e.g.,
\citealp{Furlanetto2006}), is unlikely to be resolved in observations even
with the SKA (e.g., \citealp{Mellema2013}). However, it is expected that X-ray preionization affects
the topology of the 21~cm signal in a similar manner also on much
larger scales than investigated here (e.g., \citealp{Pritchard2007}; \citealp{Baek2010}; 
\citealp{Mesinger2013}).
\par
Figure~\ref{Fig:21cm} shows images of the differential brightness
temperature in the high-resolution region and centred on the location
of the first stellar binary in the runs with and without HMXBs at
three characteristic redshifts $ z=27.8, 25.9$, and $18.1$,
assuming that $T_{\rm s}=T_{\rm gas}$. At $z = 27.8$, in both simulations, the 
photoheated relic HII region with radius $r\approx 2-3 \kpc$  
left behind by the first Pop~III binary is characterized by a weak emission 
signal $0 \mK \lesssim \delta T_{\rm b} \lesssim 50 \mK$. Higher brightness temperatures are 
achieved in dense neutral pockets and filaments of gas that survived photoevaporation 
(e.g., \citealp{Kuhlen2006}). The 21 cm signal outside the relic HII region is very different in the two simulations. 
In the simulation with HMXBs, X-ray heating induces the gas to glow in emission. In the
simulation without HMXBs, on the other hand, the gas remains cold and is observed in 
absorption. These trends are qualitatively consistent with those in previous works
\citep{Chen2008, Zaroubi2007, Thomas2008, Tokutani2009, Venkatesan2011, Yajima2013},
and the claim of a prominent 21~cm absorption feature of order
$\delta T_{\rm b} \sim -150 \mK$ as an observational signature of the first stars 
(e.g., \citealp{Cen2006}; \citealp{Chen2008}).
\par
Figure~\ref{Fig:pdf} compares the probability distribution function of
the differential brightness temperature for the two limiting cases of
collisional-only and perfect coupling at the three redshifts discussed
above. In the simulation with HMXBs, the probability distributions are
insensitive to the assumptions made about the spin temperature. At late times,
$z = 18.1$,
the distributions peak around $\delta T_{\rm b} = 0$,
because of the strong preionization of the gas by HMXBs,
confirming the visual impression of Figure~\ref{Fig:21cm}. In the
simulation without HMXBs, on the other hand, the brightness
temperature distribution is initially sensitive to the assumptions made about the spin
temperature. At $ z=27.8$ and $25.9$, assuming pure collisional
coupling implies significantly larger negative peak brightness
temperatures ($\delta T_{\rm b}\sim-50 \mK$) and 
hence a weaker 21~cm signal than assuming that the
spin temperature equals the gas temperature ($\delta T_{\rm b}\sim
-100 \mK$). At $z = 18.1$, independent of the assumptions on the spin
temperature, the differential brightness temperature distribution
peaks near $\delta T_{\rm b} \sim 50 \mK$. This contrasts with the location
of the peak around $\delta T_{\rm b} = 0$ in the simulation with
HMXBs, and may offer a route to deriving observational constrains on the nature
of the high-redshift X-ray sources (e.g., \citealp{Kuhlen2006}). 
\par

Additional constraints on X-ray preionization will
come from measurements of the optical depth towards reionization
(e.g., \citealp{Ricotti2004}) and measurements of the kinetic Sunyaev-Zeldovich (kSZ) effect
(e.g., \citealp{Zahn2012}; \citealp{Mesinger2013}; \citealp{HP2013}). 
In combination with observations of high-redshift
quasars (e.g., \citealp{Lidz2007}; \citealp{Datta2012}), these measurements will help 
uncovering the properties of the first sources of ionizing radiation and the role of 
X-rays in the early Universe.

\section{Summary and conclusions}
\label{Sec:SC}
\par
High Mass X-ray Binaries (HMXBs) are among the most luminous X-ray
sources in the local Universe. Recent high-resolution simulations with
realistic cosmological initial conditions have suggested that Pop~III
stars typically form in binaries or as members of small multiple systems, in departure
from the preceding paradigm in which the first stars formed in
isolation, thus rendering HMXBs plausible in the high-$z$ Universe as
well. Furthermore, the realization that low metallicity favours the
formation of HMXBs has led to an increased interest in the impact of
HMXBs on early cosmic history. In this paper, we carried out a set of
two radiation hydrodynamic simulations with and without HMXBs to
explore the X-ray feedback from black holes accreting diffuse gas and from
HMXBs on the surrounding gas, on the larger-scale IGM, on subsequent
star formation, and on early reionization. 
\par
It is often suggested that X-rays enhance the rate of star formation
because additional ionization boosts the abundance of $\rm H_2$, thus
promoting the cooling of the primordial gas. While our simulation with
HMXBs shows such an enhancement in $\rm H_2$, we find no strong net
effect of X-rays on star formation. The number of haloes
with cold gas fraction $\gtrsim$ 0.1 is comparable in the simulations
with and without HMXBs, and the critical mass for the onset
of cooling is insensitive to the presence of HMXBs. Furthermore, the comparison
with our preceding work in \citet{Jeon2012} suggests that the impact of HMXBs on star formation is subject to the duty
cycle of X-ray emission, such that a longer duty cycle favours a stronger
positive feedback. 
\par
We should note that we have neglected supernova feedback from Pop~III stars,
which transfers kinetic energy and propagates metals into the
surrounding medium. The metal-enriched medium, capable of forming less
massive Population~II stars, might experience a rather different
feedback history.  However, it is a matter of active debate when the
transition from Pop~III to Pop~II star formation occurred and when
Pop~III star formation terminated. Some studies have suggested that
Pop~III stars may still form until $z\sim 6$, in regions that are
uncontaminated by SN enrichment \citep[e.g.][]{Trenti2009,
  Muratov2013}. 
\par
Our simulations suggest that ionization by X-ray photons from HMXBs
can provide a significant net positive feedback on
reionization. The star formation rate is unaffected, but 
the clumping factor, which parametrizes the recombination rate in the IGM, is reduced by
a factor of $\sim 2$ by the feedback from HMXBs. Such a reduction in the
recombination rate makes it easier to keep the reionized gas ionized.  
We also examine the effect of X-ray feedback on black hole growth and show that the Bondi-Hoyle black hole 
accretion rates in the presence of HMXBs are significantly reduced. This negative feedback on
BH growth could play a role in controlling the abundance of
intermediate mass BHs, in the range of $10^4-10^5 \msun$, in SMBH
formation scenarios based on seeds left behind by the first generation
of stars \citep[e.g.][]{Tanaka2012}.
\par
Lastly, we compute the differential brightness temperature of the
neutral hydrogen 21~cm signal. Our simulations show that a promising
way for observing the signature of X-rays in the high-$z$ Universe is
through the 21~cm emission signal imprinted by X-ray heating. This
signature might depend on the characteristics of the X-ray sources,
such as their abundance, luminosity, and effective lifetime. In
addition, thanks to the pre-ionization by X-rays, the IGM around X-ray
sources can possibly achieve earlier reionization, reflected in the
location of the peak of the brightness temperature distribution.
Thus, future long-wavelength radio observations offer the exciting
prospect for constraining the nature of high-redshift X-ray sources.

\section*{acknowledgements}
We are grateful to Volker Springel, Joop Schaye, and Claudio Dalla
Vecchia for letting us use their versions of GADGET and their data
visualization and analysis tools. We further thank Volker Springel for
letting us use his implementation of the halo finder FOF as well as
his substructure finder SUBFIND. We thank Claudio Dalla Vecchia for
his implementation of additional variables in SUBFIND and valuable
discussions regarding its use. We thank Ali Rahmati, Milan Rai\v
cevi\'c, and Joop Schaye for discussions of the radiative transfer
code TRAPHIC. We thank Ali Rahmati for commenting on an earlier version 
of the draft. We thank Craig Booth for discussions of numerical
implementations of black hole growth. V.~B.\ and M.~J.\ thank the
Max-Planck-Institut f\"ur Astrophysik (MPA) for its hospitality during
part of the work on this paper. V.~B.\ and M.~M.\ acknowledge support
from NSF grant AST-1009928 and NASA ATFP grant NNX09AJ33G.
A.~H.~P.\ receives funding from the European Union's Seventh Framework
Programme (FP7/2007-2013) under grant agreement number
301096-proFeSsoR. The simulations were carried out at the Texas
Advanced Computing Center (TACC).

\bibliography{myrefs2}{}
\bibliographystyle{mn2e}

\end{document}